\documentclass[twocolumn,trackchanges]{aastex62}
\usepackage{amsmath, amssymb}
\usepackage{float}
\graphicspath{{./}{figures/}}

\def\bfd{\boldsymbol{d}}

\submitjournal{ApJ}

\shorttitle{}
\shortauthors{Kaurov et al.}
\begin{document}

\title{Highly Magnified Stars in Lensing Clusters: New Evidence in a Galaxy Lensed by MACS J0416.1-2403}

\correspondingauthor{Alexander A. Kaurov}
\author{Alexander A. Kaurov}
\author{Liang Dai}
\author{Tejaswi Venumadhav}
\affiliation{Institute for Advanced Study, 1 Einstein Drive, Princeton, NJ 08540, USA}
\author{Jordi Miralda-Escud\'{e}}
\affiliation{Instituci\'o Catalana de Recerca i Estudis Avan\c cats, Barcelona, Catalonia}
\affiliation{Institut de Ci\`encies del Cosmos, Universitat de Barcelona (IEEC-UB), Barcelona, Catalonia}
\affiliation{Institute for Advanced Study, 1 Einstein Drive, Princeton, NJ 08540, USA}
\author{Brenda Frye}
\affiliation{Department of Astronomy/Steward Observatory, University of Arizona, 933 N. Cherry Ave, Tucson, AZ  85721,  USA}
\email{kaurov@ias.edu}

\begin{abstract}

We examine a caustic-straddling arc at $z=0.9397$ in the field of the galaxy cluster MACS J0416.1-2403 ($z=0.397$) using archival multiband HST images and show that its surface brightness exhibits anomalies that can be explained by a single highly magnified star undergoing microlensing. First, we show that the surface brightness pattern is not perfectly symmetric across the cluster critical curve, which is inconsistent with a locally smooth lens model; the location of the candidate star exhibits the most significant asymmetry. Second, our analysis indicates that the asymmetric feature has $\sim 30\%$ higher flux in the 2012 visits compared to the Frontier Fields program visits in 2014. Moreover, the variable asymmetric feature shows an anomalous color between the F814W and F105W filters in 2014. These anomalies are naturally explained by microlensing induced variability of a caustic-transiting blue supergiant in a star-forming region, with a mean magnification factor around $\mu \sim 200$. We extend this study to a statistical analysis of the whole arc image and find tentative evidence of the increased mismatch of the two images in the proximity of the critical line. Robust detection of one or multiple caustic-transiting stars in this arc will enable detailed follow-up studies that can shed light on the small-scale structure of the dark matter inside the cluster halo.

\end{abstract}

\keywords{gravitational lensing: micro --- galaxies: clusters: individual (MACS J0416.1-2403) }

%%%%%%%%%%%%%%%%%%%%%%%%%%%%%%%%%%%%%%%%%%%%%%%%%%%%%%%%%%%%%%%
\section{Introduction} 
\label{sec:intro}
%%%%%%%%%%%%%%%%%%%%%%%%%%%%%%%%%%%%%%%%%%%%%%%%%%%%%%%%%%%%%%%

Galaxy clusters are Nature's most powerful gravitational lenses. Bright stars in the background that move across a cluster lensing caustic reach huge magnification factors and can be individually detectable by optical or infrared telescopes~\citep{1991ApJ...379...94M}. The most luminous of these stars are typically part of star-forming galaxies straddling a caustic, with transverse motions dominated by the bulk flow of the large-scale structure. The first example of these caustic transiting stars was a highly magnified blue supergiant in a lensed spiral galaxy at $z \simeq 1.5$ behind the galaxy cluster MACS J1149, which was discovered in May 2016 during the Hubble Space Telescope (HST) Frontier Fields program (FFP)~\citep{2018NatAs...2..334K}. The second example has recently been reported in the lensing cluster MACS J0416~\citep{Chen:2019ncy}. Two other possible candidates in the same cluster were reported by \cite{2018NatAs...2..324R}.
Future observations of caustic transients promise to offer great insight into massive stars in high-redshift galaxies~\citep{2018NatAs...2..334K}, Population III stars in the early universe~\citep{windhorst2018observability}, or other cosmological sources~\citep{diego2018universe}.

In the caustic vicinity, lensing of compact sources is highly sensitive to small-scale granularity in the lens mass distribution. This offers new possibilities to explore small-scale structures in the dark matter (DM) in cluster halos using caustic transiting stars, which are not generally detectable in other observations. 

Intracluster stars are inevitably present in lensing clusters, so the impact of any DM substructure can only be demonstrated if it is distinguished from the effects of microlensing by known stars~\citep{2017ApJ...850...49V, 2018ApJ...857...25D, Oguri:2017ock}. Microlensing can make a highly magnified star intermittently detectable, as it occasionally boosts the magnification factor above the value without microlensing at a given source position. In the case of MACS J1149, microlensing has yielded the best constraints on compact DM in the mass range $0.1$--$10\,M_\odot$~\citep{Oguri:2017ock}. Future observations have the potential to enable constraints across a wide mass range ~\citep{2017ApJ...850...49V, 2018ApJ...857...25D} that are more stringent than those from Galactic and Local group microlensing surveys~\citep{alcock2001macho, tisserand2007limits, griest2013new, niikura2017microlensing}

Close to the critical curve, DM substructure in the cluster halo can induce enhanced astrometric perturbations~\citep{minor2017robust} that render the surface brightness pattern asymmetric across the critical curve~\citep{2018ApJ...867...24D}. Detecting multiple caustic transiting stars in a lensed galaxy enables astrometric asymmetry measurements that can probe the cluster DM subhalo content in the mass range $\sim 10^6$--$10^8\,M_\odot$~\citep{2018ApJ...867...24D}.

To fully realize the scientific potential of caustic transients to probe the nature of the DM, detecting many highly magnified stars is necessary. In this study, we report on our search of caustic transiting stars in archival HST data. We find a candidate in a lensed galaxy in the field of the FFP galaxy-cluster MACS J0416. We present multiple pieces of evidence for the highly magnified star interpretation: the asymmetry in the surface brightness across the lensing critical curve (presented in \S\ref{sec:flux}), temporal flux variability and an anomalous color between filters at the same position where flux asymmetry and variability are indicated by the observations (\S\ref{sec:slitF}). In \S4 we introduce the microlensing simulations,and
in \S5 we discuss the properties of the source star.  In \S\ref{sec:fits}, we explore additional possibilities of statistical analysis of surface brightness mismatch across a critical curve. Final remarks will be given in \S\ref{sec:discussion}.

As this work was being finished, a similar effort by \cite{Chen:2019ncy} was posted in a preprint, in which they detect the same highly magnified star. We comment in our discussion section on the difference between our analysis and theirs, and find that we agree on most of the conclusions. Remarkably, two independent searches in the Frontier Field clusters have identified the same candidate super-magnified star, even though the specific microlensing variations we detected were on different epochs. This strengthens our confidence on the reality of this new caustic-transiting star.

%%%%%%%%%%%%%%%%%%%%%%%%%%%%%%%%%%%%%%%%%%%%%%%%%%%%%%%%%%%%%%%%%%%%%%%%%%%%%%%%
\begin{figure}
    \centering
    \includegraphics[width=\columnwidth]{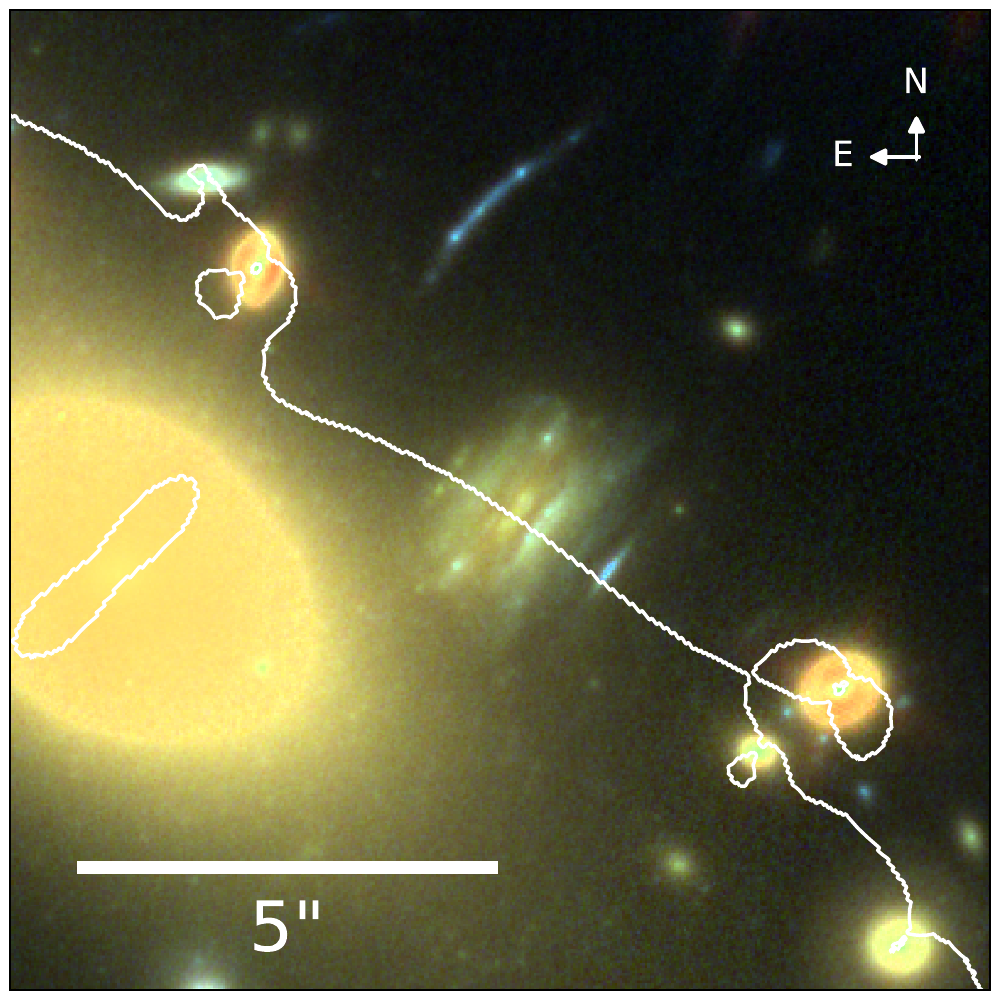}
    \caption{Combined false-color image using the F105W, F814W and F606W filters showing the $12\arcsec \times 1 2\arcsec$ region centered on the set of lensed images of interest in MACS J0416.1-2403. The white line is the lensing critical curve for a source redshift $z=0.94$ \citep{2017A&A...600A..90C}.}
    % (64d02m12.04s -24d04m02.04s)
    \label{fig:environment}
\end{figure}
%%%%%%%%%%%%%%%%%%%%%%%%%%%%%%%%%%%%%%%%%%%%%%%%%%%%%%%%%%%%%%%%%%%%%%%%%%%%%%%%

%%%%%%%%%%%%%%%%%%%%%%%%%%%%%%%%%%%%%%%%%%%%%%%%%%%%%%%%%%%%%%%%%%%%%%%%%%%%%%%%%
\section{Flux asymmetry across the critical line}
\label{sec:flux}
%%%%%%%%%%%%%%%%%%%%%%%%%%%%%%%%%%%%%%%%%%%%%%%%%%%%%%%%%%%%%%%%%%%%%%%%%%%%%%%%%

%%%%%%%%%%%%%%%%%%%%%%%%%%%%%%%%%%%%%%%%%%%%%%%%%%%%%%%%%%%%%%%%%%%%%%%%%%%%%%%%%
\begin{figure*}
    \centering
    \hspace{15pt}\includegraphics[width=0.28\textwidth]{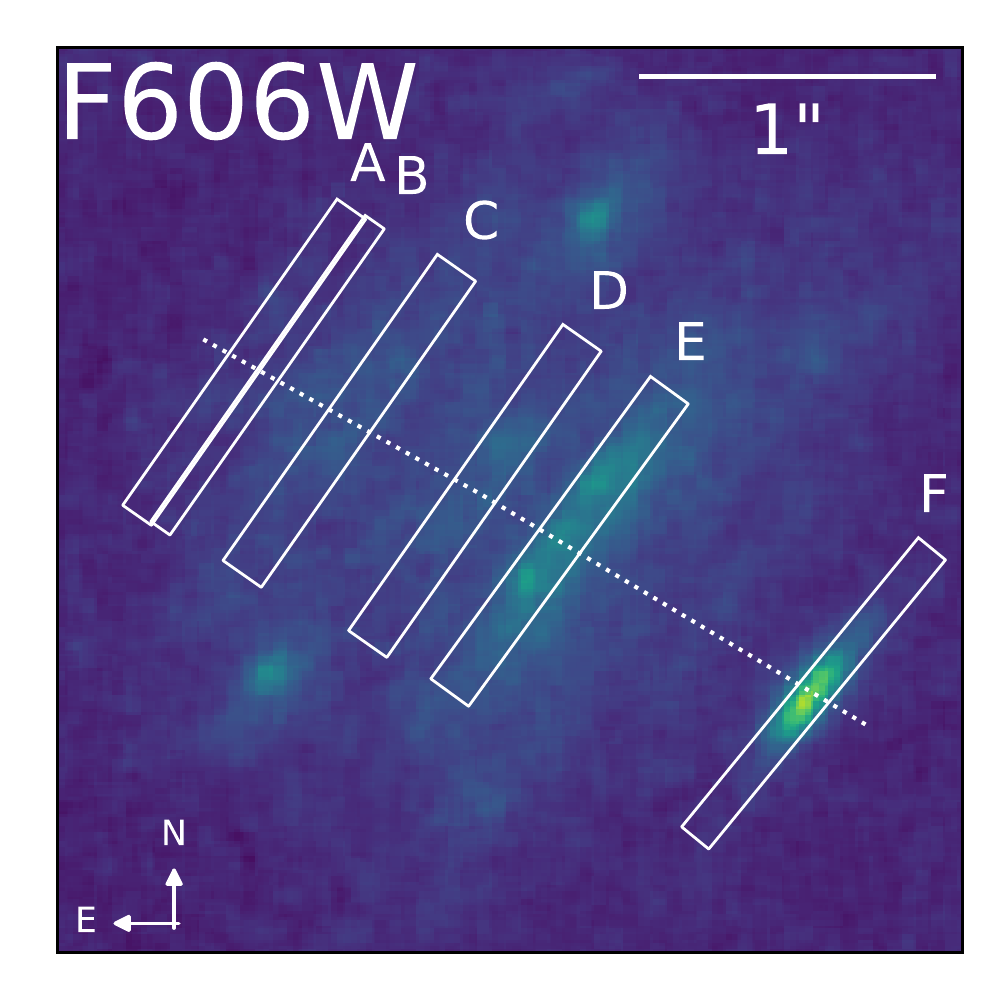}
    \hspace{10pt}\includegraphics[width=0.28\textwidth]{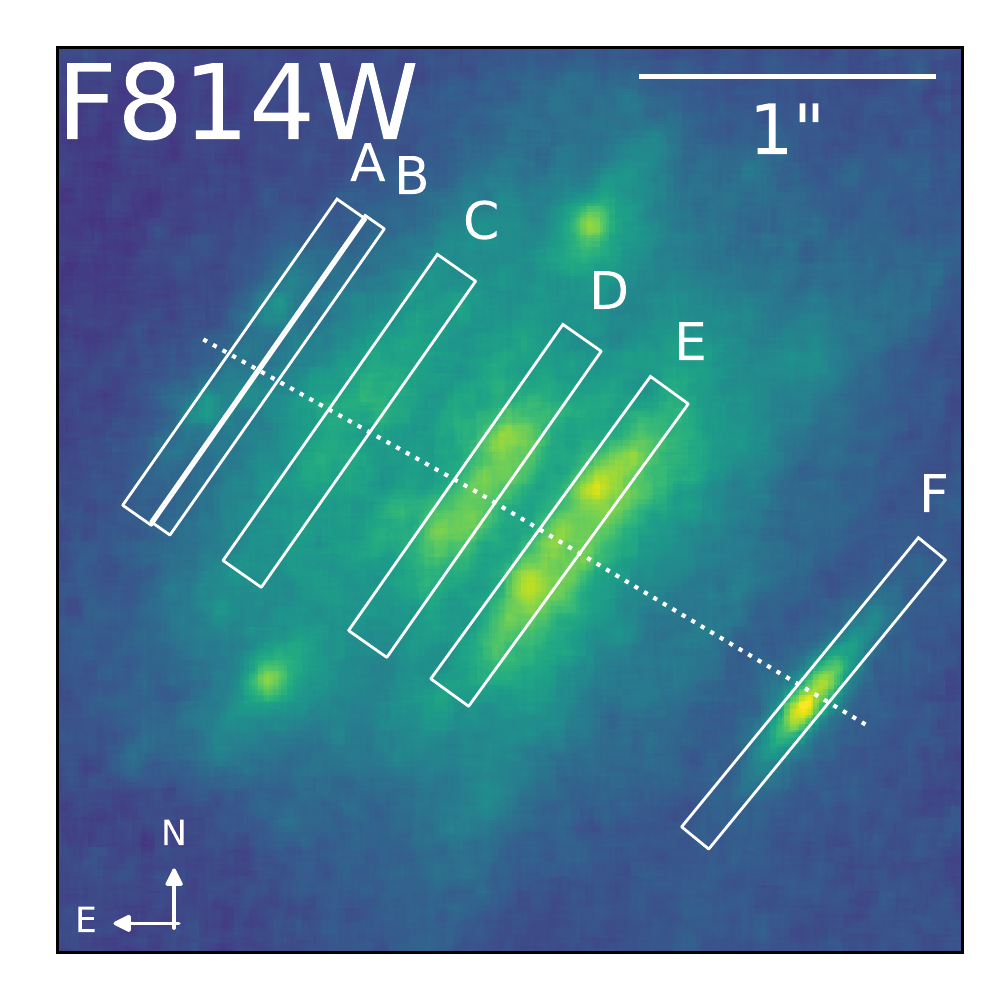}
    \hspace{10pt}\includegraphics[width=0.28\textwidth]{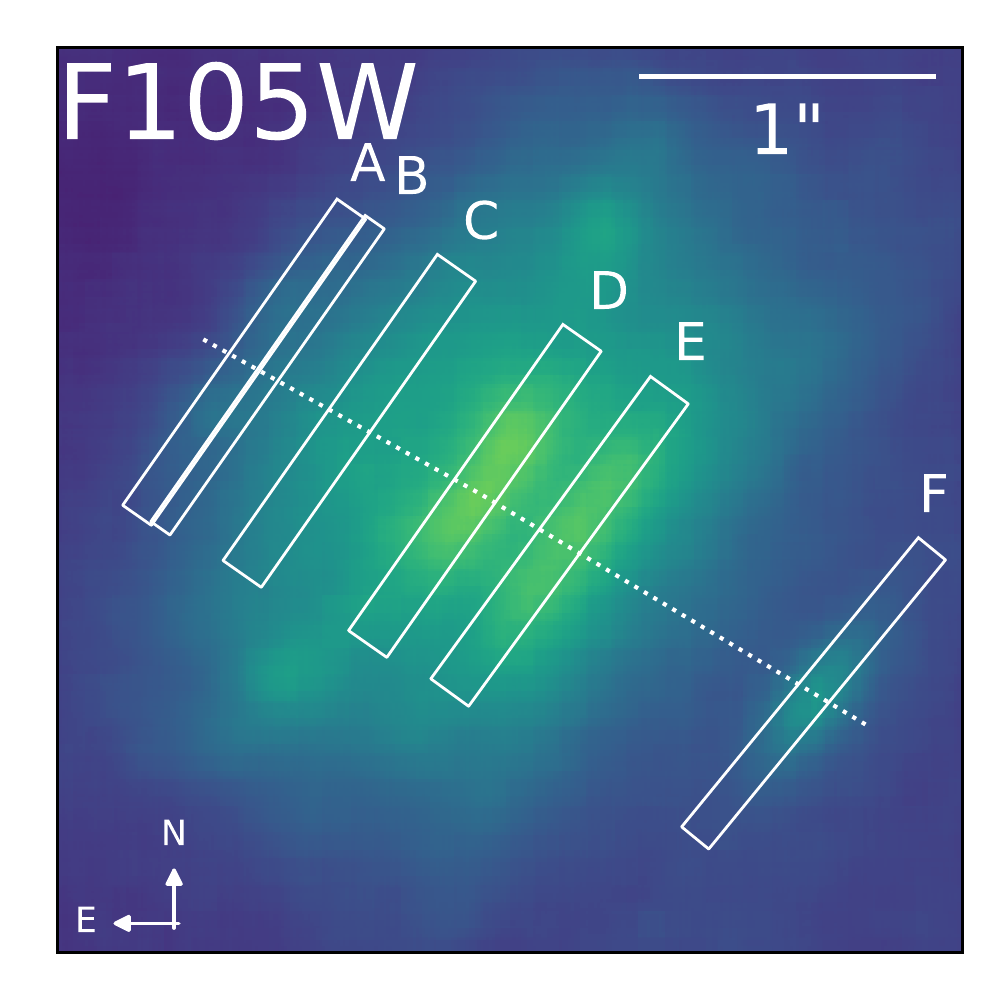}\\
    \includegraphics[width=0.3\textwidth]{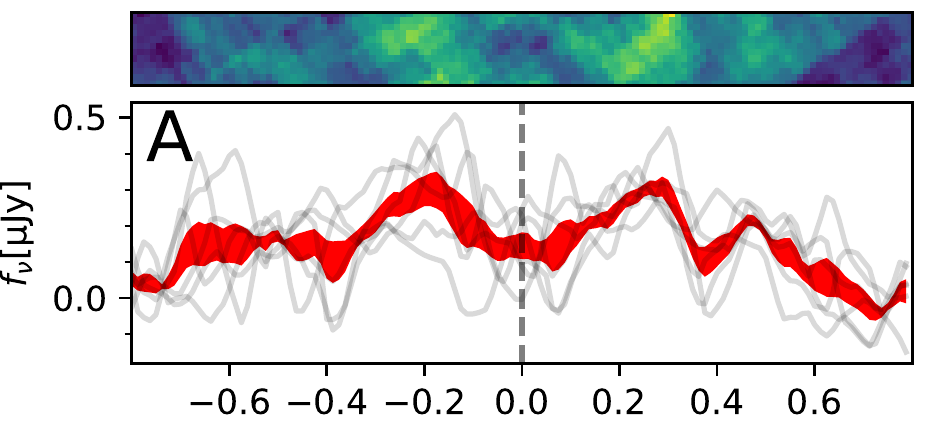}
    \includegraphics[width=0.3\textwidth]{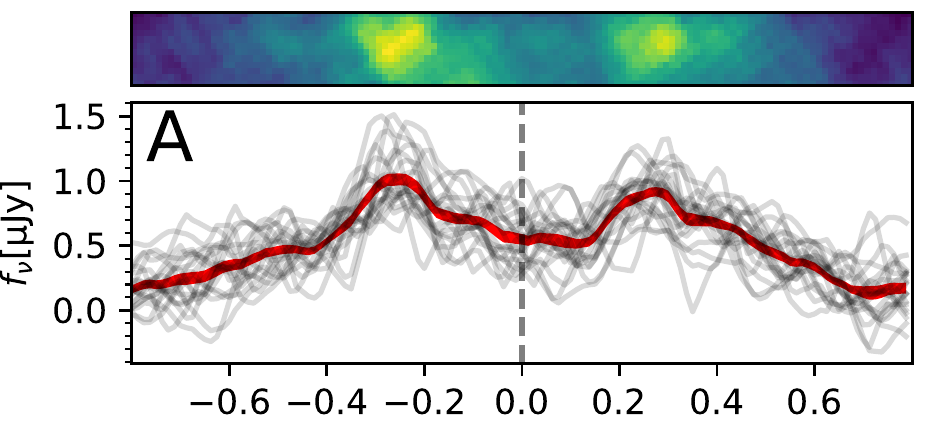}
    \includegraphics[width=0.3\textwidth]{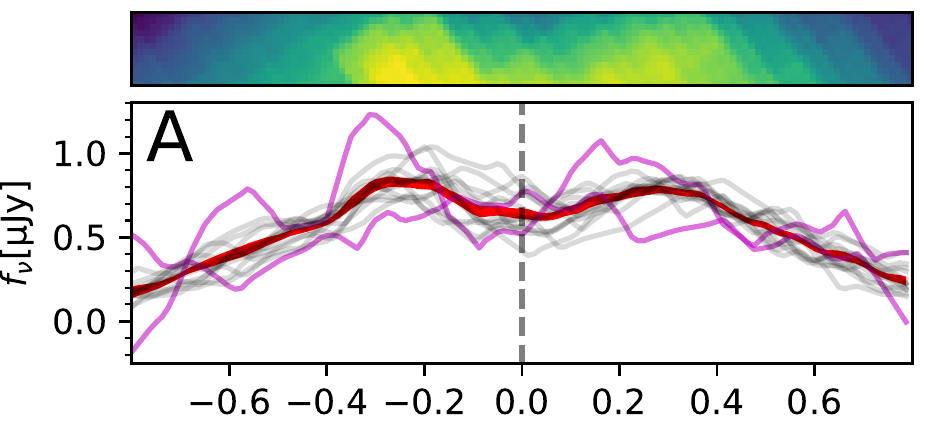}\\
    \includegraphics[width=0.3\textwidth]{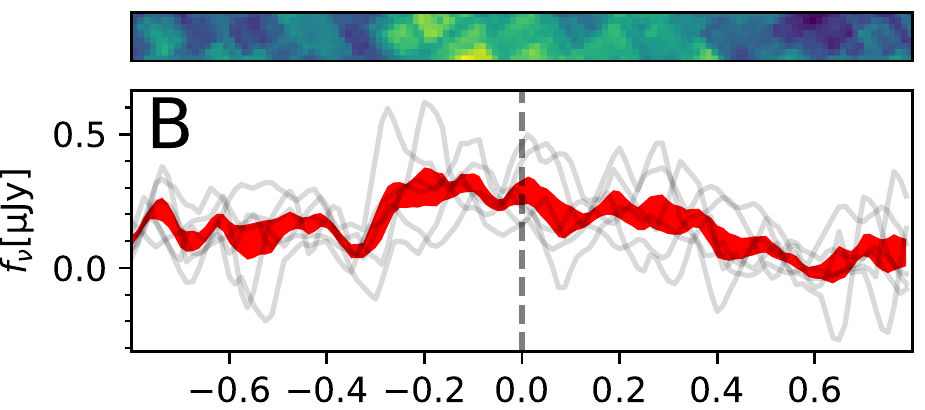}
    \includegraphics[width=0.3\textwidth]{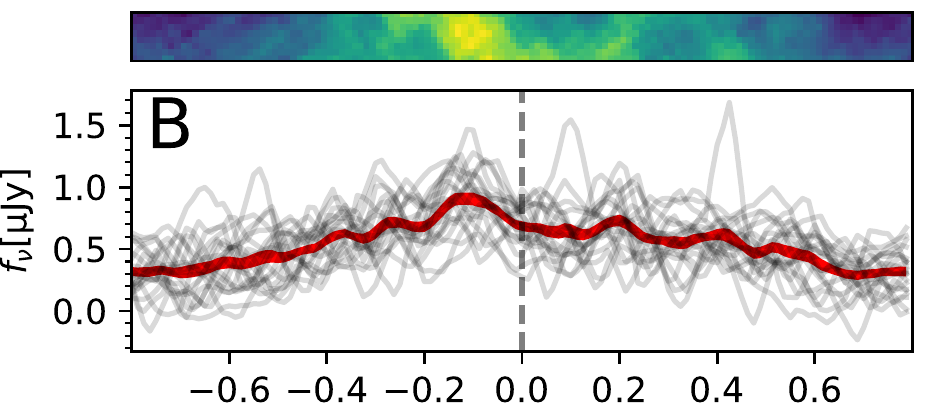}
    \includegraphics[width=0.3\textwidth]{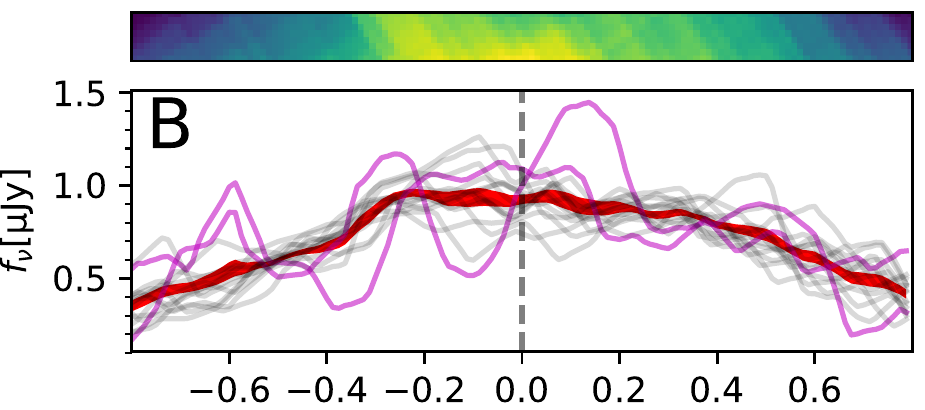}\\
    \includegraphics[width=0.3\textwidth]{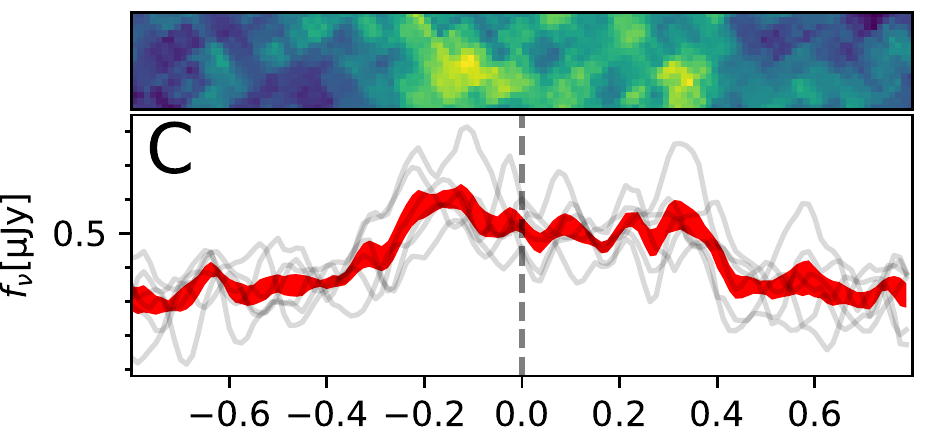}
    \includegraphics[width=0.3\textwidth]{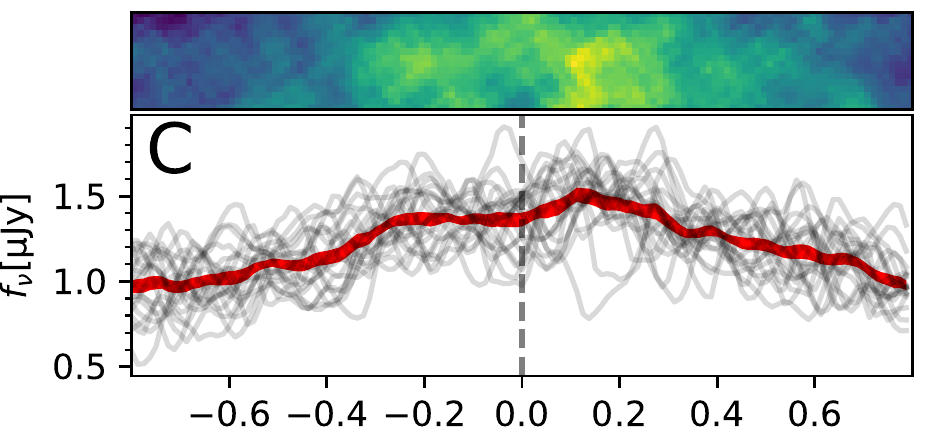}
    \includegraphics[width=0.3\textwidth]{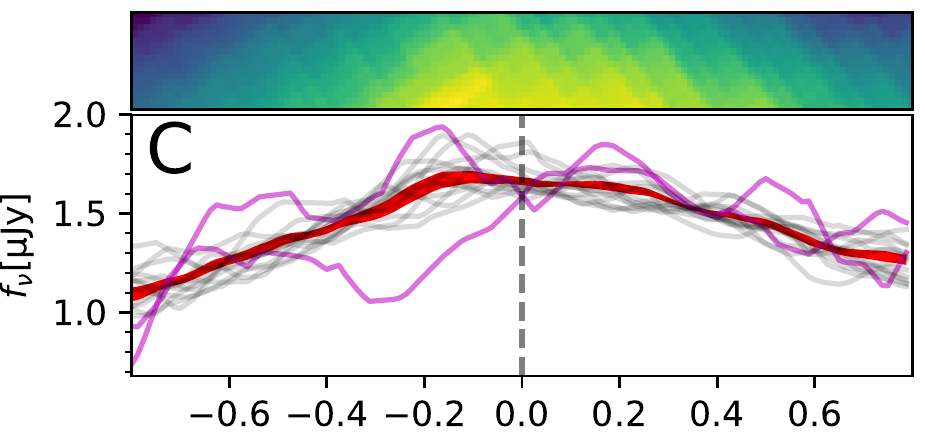}\\
    \includegraphics[width=0.3\textwidth]{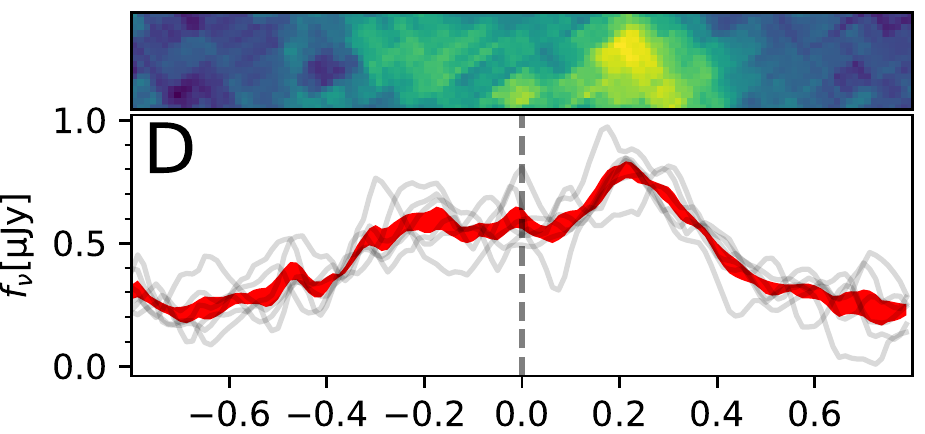}
    \includegraphics[width=0.3\textwidth]{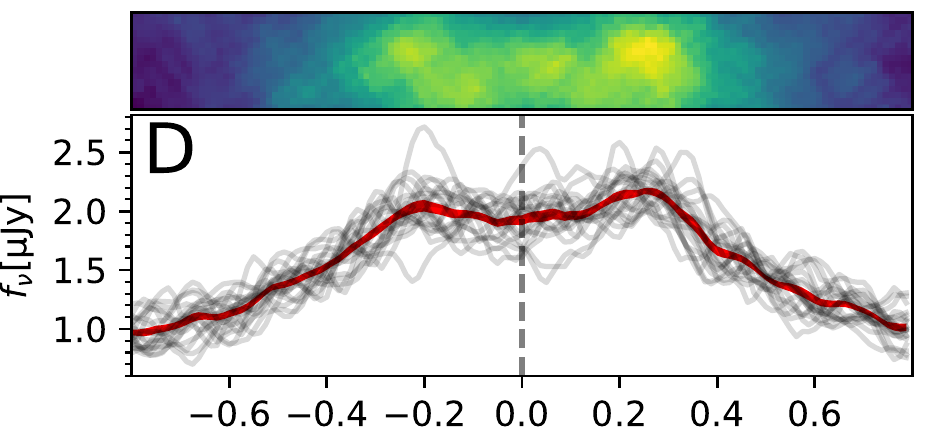}
    \includegraphics[width=0.3\textwidth]{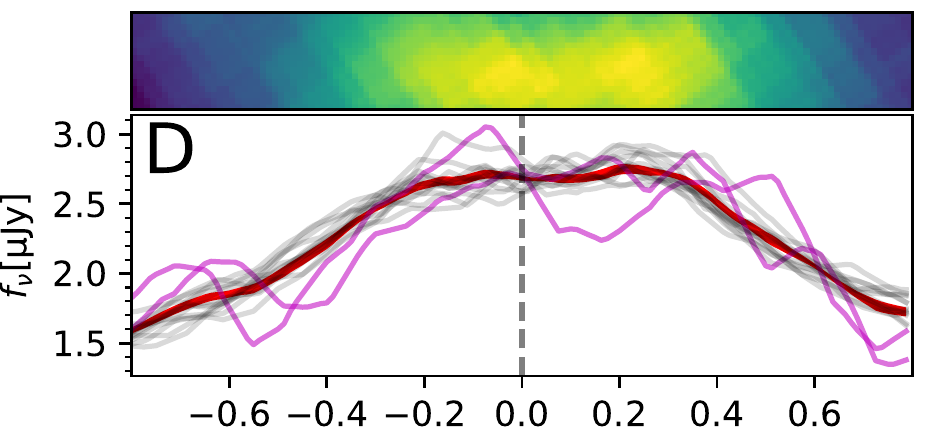}\\
    \includegraphics[width=0.3\textwidth]{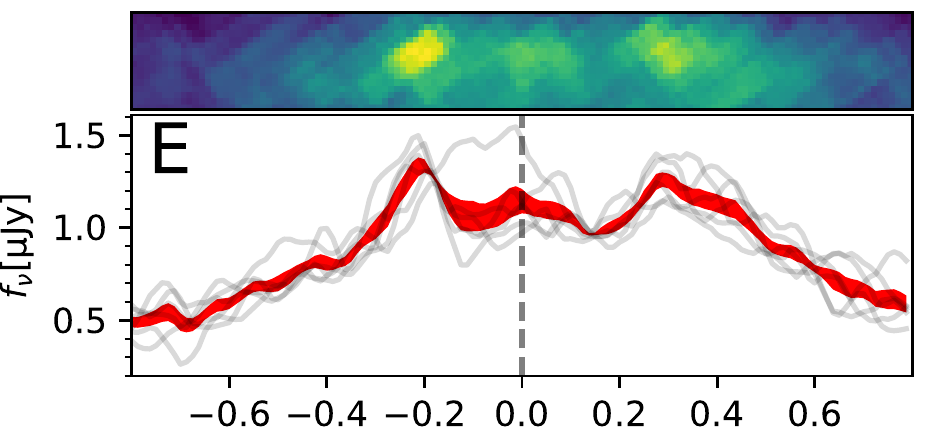}
    \includegraphics[width=0.3\textwidth]{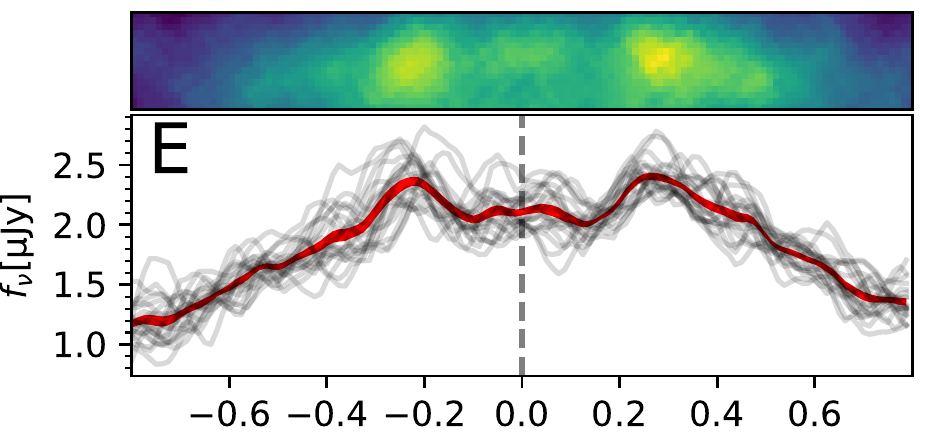}
    \includegraphics[width=0.3\textwidth]{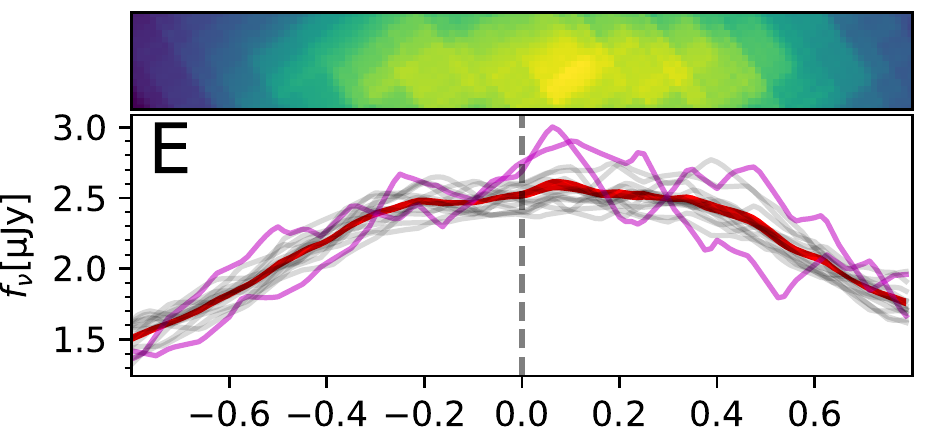}\\
    \includegraphics[width=0.3\textwidth]{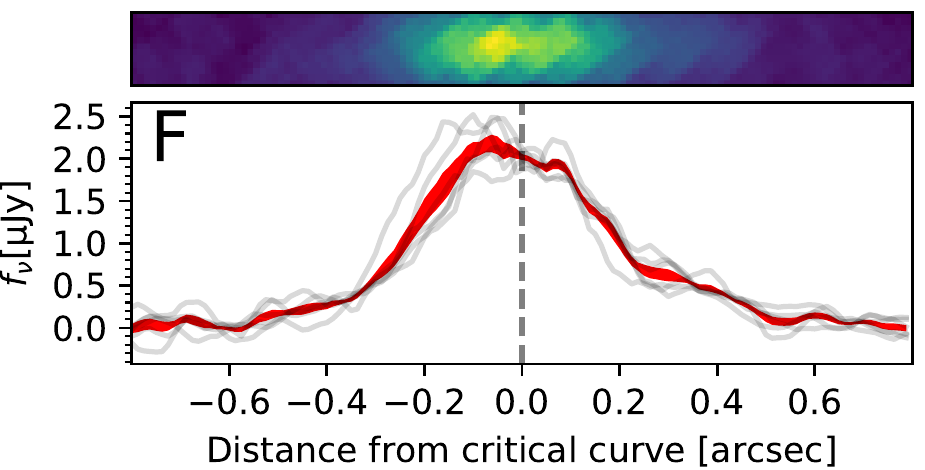}
    \includegraphics[width=0.3\textwidth]{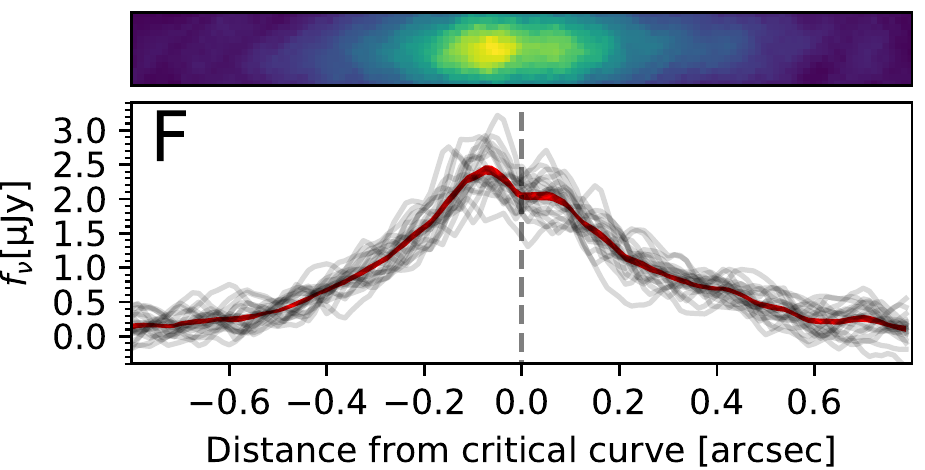}
    \includegraphics[width=0.3\textwidth]{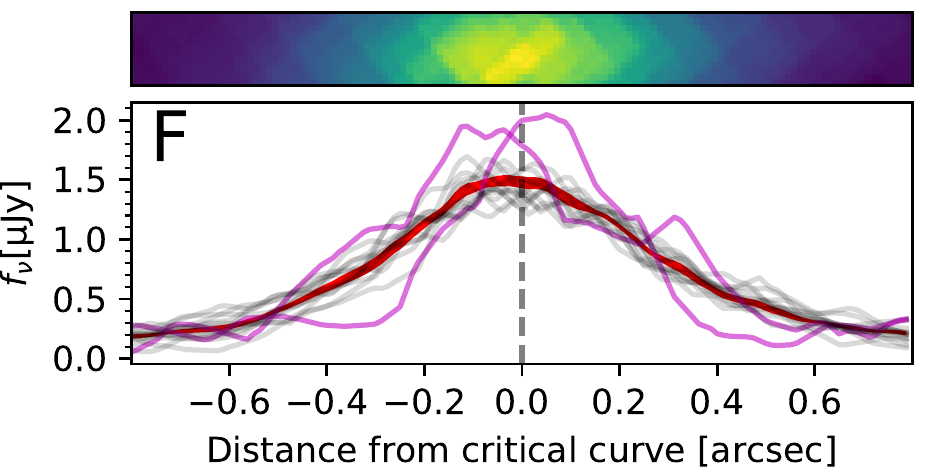}
    \caption{Top panels show $3.5\arcsec\times3.5\arcsec$ cutouts in the F606W, F814W and F105W filters and 6 slits. Lower panels show brightness profiles along each slit. Gray curves show individual visits (each combining 4 exposures) from 2014 that are part of the Frontier Fields Program (FFP). The two magenta curves on the rightmost panels are for two visits in 2012 (see \S\ref{sec:temp}) with lower SNRs compared to the gray curves (1.5 and 1.3 ks of exposure respectively; while the gray curves correspond to 5.6 ks of exposure each). Red bands show the statistical uncertainty (1$\sigma$) around the mean values. Vertical dashed lines mark the expected position of the critical curve. The most significant asymmetries between the interior and exterior of the critical curve are found in slit F in F814W and F606W, and also in slit D in F606W.}
    \label{fig:comparison}
\end{figure*}
%%%%%%%%%%%%%%%%%%%%%%%%%%%%%%%%%%%%%%%%%%%%%%%%%%%%%%%%%%%%%%%%%%%%%%%%%%%%%%%%%

In this section we adopt Hubble Legacy Archive (HLA) products of data obtained by the FFP (PI: Lotz, GO-13496) in 2014. Deep images are available in 3 ACS filters (F435W, F606W, F814W) and 4 WCF3 filters (F105W, F125W, F140W, F160W). The images of individual visits are already drizzled and cleaned of defects such as cosmic ray hits; however, they are not aligned, and therefore we align them ourselves.  We focus this present study on a group of $\sim$10 images that do not appear highly elongated, probably because the source happens to be intrinsically elongated in a direction almost parallel to the caustic. Despite this, we refer to this arc structure as the ``arc''.  A co-added false color image, with this arc at the center, is shown in Figure \ref{fig:environment}.

To investigate the asymmetry of the arc, we first subtract the intracluster light (ICL). We use the following smoothly-varying ICL model. We define a $R = 6\arcsec$ annulus with $2\arcsec$ width that surrounds the arc but does not contain any sources except the ICL. Then, we fit a second-order two-dimensional polynomial to the fluxes in pixels on the annulus. Extrapolation of the best-fit to inside the annulus gives the ICL model for the arc. We tested other ICL models, including an elliptical profile fit around the nearest BCG, a local gradient model, or no ICL subtraction at all. We found that our results are not sensitive to the choice of the ICL model, for the reason that the arc is relatively compact and therefore any two pixels lying close to the critical curve have similar ICL contributions.

We search the co-added images for any flux difference between pairs of pixels that lie symmetrically across the critical curve.  As shown in Figure \ref{fig:comparison}, the direction of arc elongation is almost perpendicular to the critical curve. To highlight possible flux asymmetries within this large arc structure, we define six individual narrow slits labeled from A to F. In the subsequent panels we show the one-dimensional distribution of the flux along each slit, which is summed along the perpendicular direction and is averaged over individual visits. The position of the critical curve is indicated by the vertical dashes at position $0\arcsec$. We show the F606W and F814W images, which have higher resolutions than the IR images and higher signal-to-noise ratios than the F435W image; we also show the F105W image, for reasons to be discussed in \S\ref{sec:slitF}. 

For slits A and E, the flux is approximately symmetric about the critical curve. The arc in slit F, however, exhibits asymmetry close to the critical curve in F814W. The red curve with a finite thickness corresponds to the 1$\sigma$ statistical uncertainty in the flux averaged over individual visits, which are themselves shown as gray lines. We estimate that the significance of the asymmetry in slit F exceeds $5\sigma$. An asymmetry of the same shape and at the same location is also confirmed in F606W. In addition, we find flux asymmetries in other slits but only in one filter --  B and C in F814W and D in F606W. These anomalies are not as strong as the one in slit F but are still significant, and they imply fainter underlying structures that are perhaps affected by microlensing.

%%%%%%%%%%%%%%%%%%%%%%%%%%%%%%%%%%%%%%%%%%%%%%%%%%%%%%%%%%%%%%%%%%%%%%%%%%%%%%%%%
\begin{figure*}
    \centering
    \includegraphics[width=\textwidth]{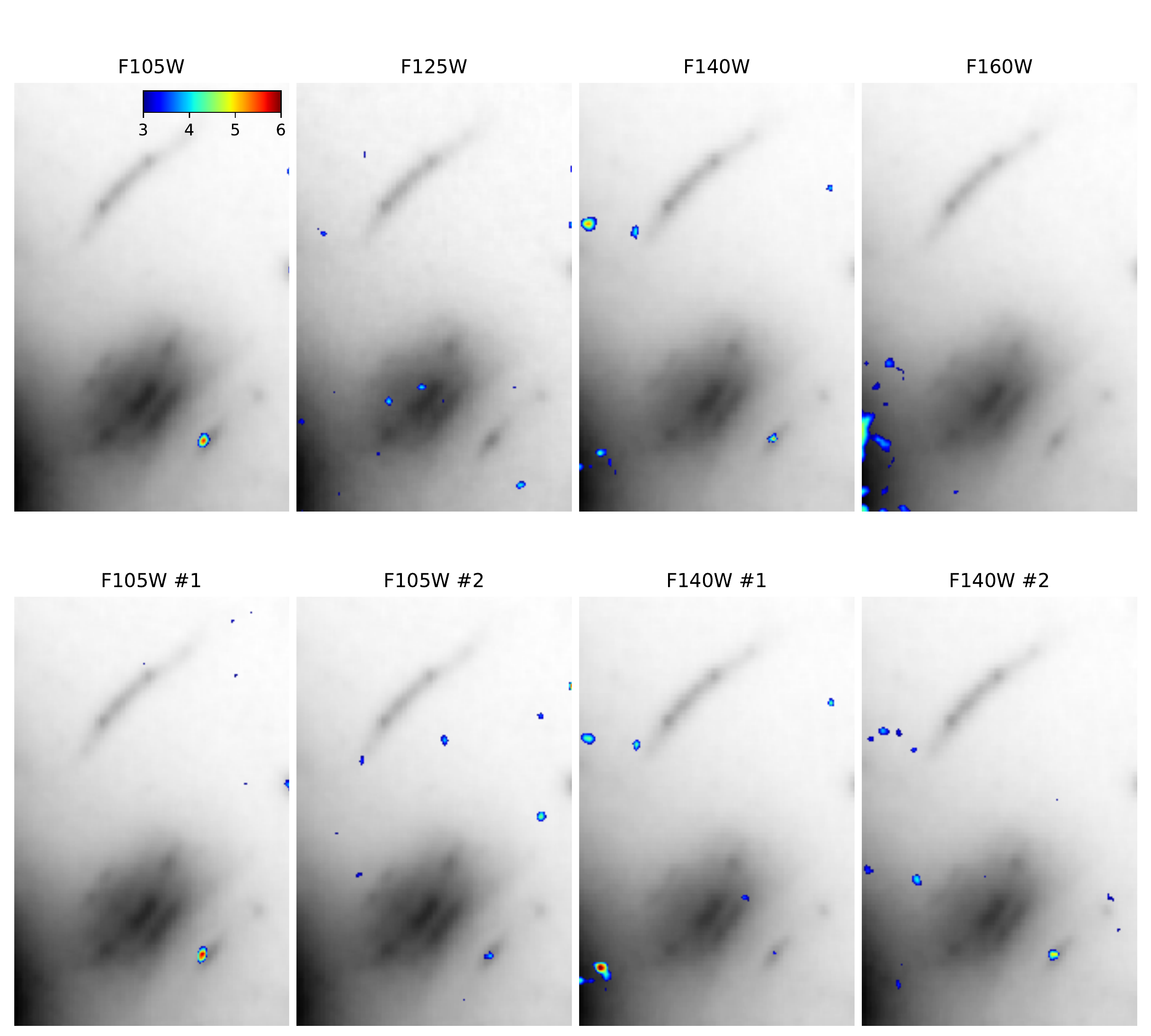}
    \caption{Difference between the 2012 observations and stacked FFP observations from 2014 in F105W, F125W, F140W and F160W. Background gray scale image shows stacked FFP visits in corresponding filters and color highlights the regions with residuals exceeding 3$\sigma$. The significance is estimated from the variance of the residual map. \textit{Upper panels:} Comparison of 2012 stacked observations in 4 filters. \textit{Lower panels:} We split F105W and F140W data into two separate visits.}
    \label{fig:dif}
\end{figure*}
%%%%%%%%%%%%%%%%%%%%%%%%%%%%%%%%%%%%%%%%%%%%%%%%%%%%%%%%%%%%%%%%%%%%%%%%%%%%%%%%%

Flux asymmetry in pixel pairs located symmetrically about the critical curve can be caused by two images of a highly magnified star or other surface brightness features in the lensed galaxy. If microlensing by cluster stars magnifies one image of a lensed star more strongly than the other, temporal variability is expected. But if asymmetric magnification is due to lensing by larger mass structures such as DM subhalos, a persistent pattern is expected when the variability timescale is too long to be observed.

F814W is the optimal filter to study these anomalies because the combination of signal-to-noise and resolution is best in this filter to probe compact sources near the critical curve. The same flux asymmetry in slit F is visible in the simultaneously acquired F606W images. This asymmetry is unlikely to be caused by a random foreground object near the critical curve because spectroscopic studies of the arc in slit F using VLT/MUSE revealed no feature at a redshift different than that of the arc, $z = 0.94$~\citep{2017A&A...600A..90C}.

In \S\ref{sec:fits} we explore a statistical method to evaluate flux asymmetry across the critical curve, which is an alternative to the above analysis using slits.

%%%%%%%%%%%%%%%%%%%%%%%%%%%%%%%%%%%%%%%%%%%%%%%%%%%%%%%%%%%%%%%%%%%%%%%%%%%%%%%%
%\newpage
\section{Detailed analysis of slit F}
\label{sec:slitF}
\subsection{Temporal variability}
\label{sec:temp}
%%%%%%%%%%%%%%%%%%%%%%%%%%%%%%%%%%%%%%%%%%%%%%%%%%%%%%%%%%%%%%%%%%%%%%%%%%%%%%%%

We examine temporal variability in the archival images in wide IR filters (F105W, F125W, F140W, F160W) between the combined visits in 2012 (PI:  Postman, GO-12459) and the combined FFP visits in 2014 (PI: Lotz, GO-13496). We calculate the difference between the two epochs and the variance in the residual map, which we use to estimate the significance of residuals at any position. The combined 2014 images are much deeper, so the errorbar on the flux difference is dominated by the 2012 measurements. The top panels of Figure \ref{fig:dif} show the significance of the residuals in all 4 filters. We see significant residuals in F105W and F140W exactly at the position in slit F where the most prominent flux asymmetry has been found. The lower panels of Figure \ref{fig:dif} show residuals between each of the two individual visits in 2012 and the combined 2014 FFP visits. In both F105W and F140W, each of the 2012 visits shows a significant flux difference from the combined FFP visits, disfavoring systematic error explanations. Further evidence against cosmic rays or other systematics is shown in the left panels of Figure \ref{fig:variability}, where a square cutout of the arc in the F105W image is shown from the stacked 2014 FFP images and from three separate exposures in the August 5th visit in 2012. The increased brightness in stack F relative to the 2014 data is seen in all individual images.

We quantify the variability applying aperture photometry to individual exposures. The right panel of Figure \ref{fig:variability} shows the flux in F105W (normalized to the value in the stacked FFP images) within a circular aperture twice the size of the PSF at multiple epochs. The fluxes in both 2012 visits are significantly higher than that of the averaged 2014 FFP visits. The significance is $>4\sigma$ and consistent with what we found in the residual maps. Notably, no other region in the cutout in the right panel of Figure \ref{fig:variability} shows flux variability at a similar level. 

The higher flux in 2012 compared to the FFP values is also seen in F140W, but is at a lower significance in F125W and F160W, which were taken approximately at the same time as F105W and F140W. We believe this is attributable to random noise.

We do not see any detectable variability in the UV/optical filters in the FFP images. The FFP had deep visits during January to February 2014 in F425W, F606W and F814W from which we detect flux asymmetry, and then another $2\,$ks exposure in September 2014 which is too shallow to be useful for our purpose. There were also visits in 2012 (PI: Postman, \#12459), but three out of four visits happened to position this arc in between the detectors, and the only one that did not miss the arc had a $1\,$ks long exposure and is too shallow.

As an aside, we point out the possible existence of another microlensed source in the thin arc at $z_{phot}\sim2.4$ \citep{2013ApJ...762L..30Z} located $\sim 4"$ North of the $z = 0.94$ arc that we focus on in this paper. We notice a compact feature which appears in between the two brightest blobs, and does not to have a counter image
(see Figure \ref{fig:environment}). This can be explained either by assuming this feature is directly located on top of the critical curve in order to preserve symmetry, or assuming it is to the southeast of the critical curve, in which case symmetry needs to be broken by microlensing. The latter case is favored by the off-center position of this feature relative to the two brightest blobs in the arc. For the feature in the middle we find a $\sigma\sim3$ brightening event during a FFP visit on January 21 of 2014 in the F814W filter, with no evidence of cosmic ray hits as judged from individual exposures. We note that this may be similar in some aspects to the fast point-like transients previously identified in a different caustic straddling arc at $z = 1$ in MACS J0416~\citep{2018NatAs...2..324R}. While we cannot claim definite evidence that this is a new microlensed object, the presence of these anomalies indicates the potential for further discoveries of microlensing in deeper images of these lensing clusters.

%%%%%%%%%%%%%%%%%%%%%%%%%%%%%%%%%%%%%%%%%%%%%%%%%%%%%%%%%%%%%%%%%%%%%%%%%%%%%%%
\subsection{Color anomaly}
\label{sec:color}
%%%%%%%%%%%%%%%%%%%%%%%%%%%%%%%%%%%%%%%%%%%%%%%%%%%%%%%%%%%%%%%%%%%%%%%%%%%%%%%

\begin{figure*}
    \centering
    \raisebox{0\height}{\includegraphics[width=0.44\textwidth]{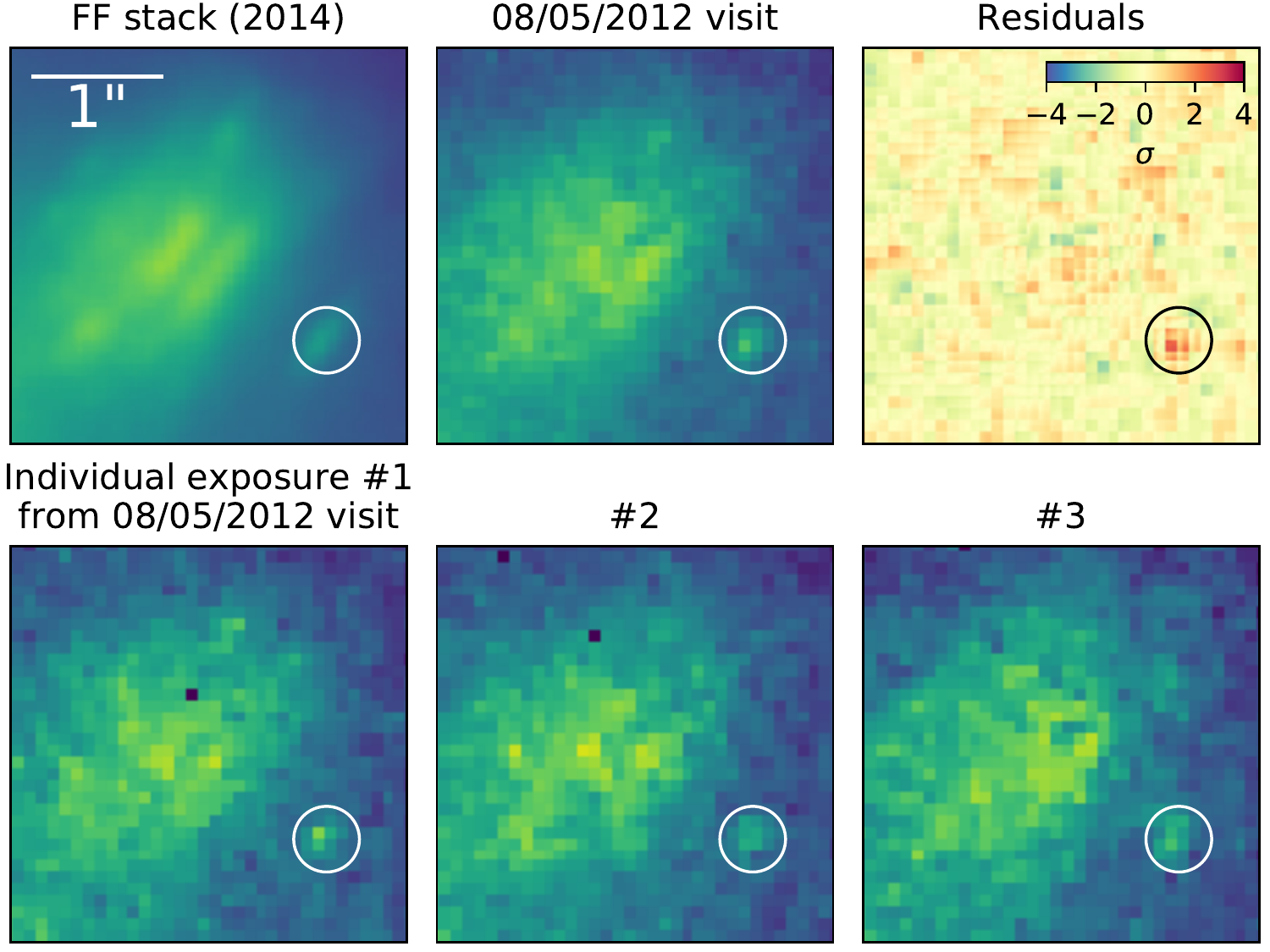}}
    \raisebox{-0.1\height}{\includegraphics[width=0.55\textwidth]{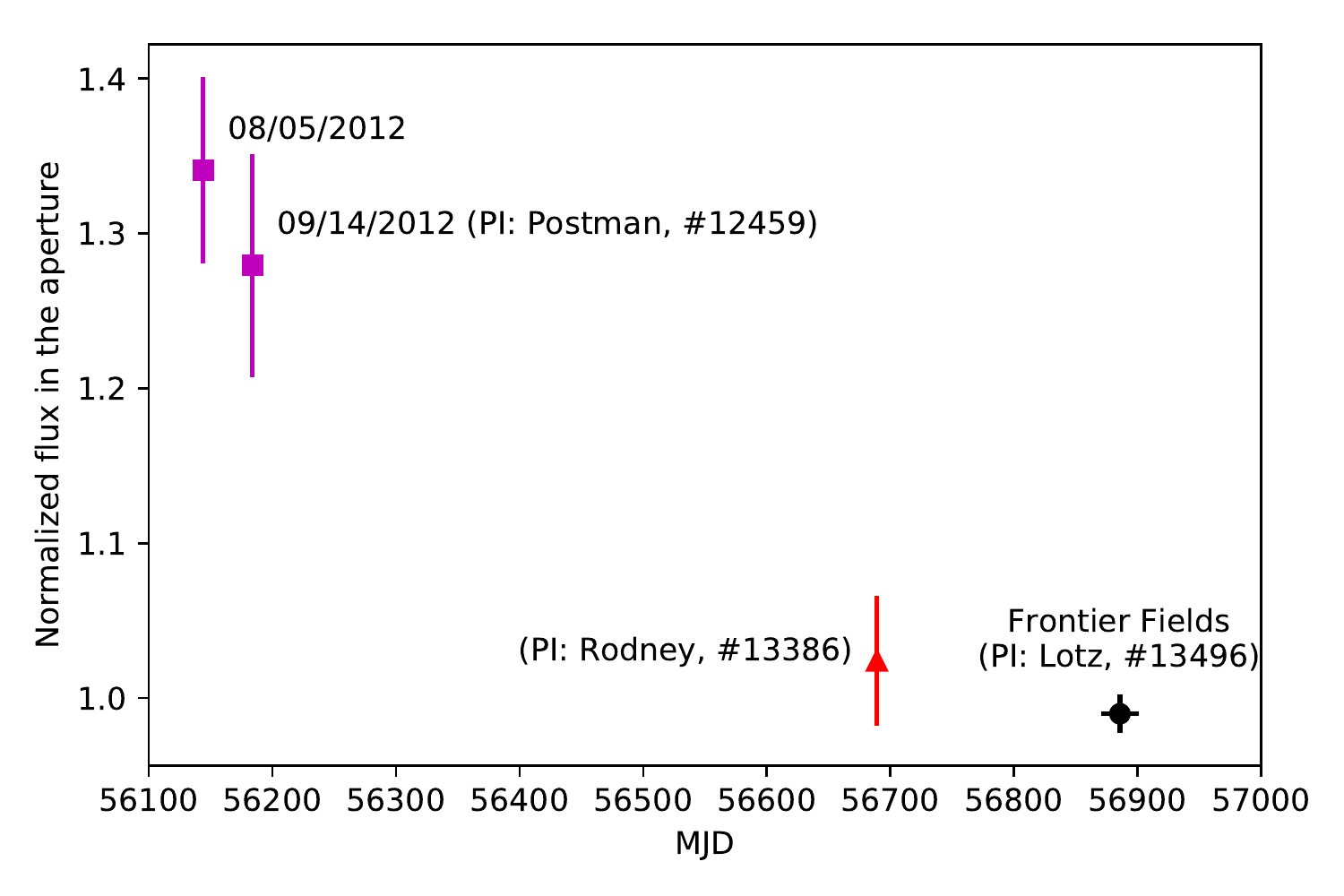}}
    \caption{\textit{Left panels, top row:} $3.1\arcsec\times3.1\arcsec$ cutouts of stacked images taken in the 2014 Frontier Fields program in F105W; image from the individual 1.5ks visit on 08/05/2012; subtraction of previous two images in units of rms noise.  \textit{Bottom row:} three 500s exposures from that visit showing consistent brightness at the circled position of the putative microlensed star -- no evidence for a cosmic ray hit.
    \textit{Right panel:} Flux variation within aperture in F105W filter. Flux is normalized to the stacked FFP images. Uncertainty is estimated from variations in individual exposures. Flux in 2012 visits is $\sim30\%$ higher than in 2014.}
    \label{fig:variability}
\end{figure*}

\begin{figure}
    \centering
    \includegraphics[width=\columnwidth]{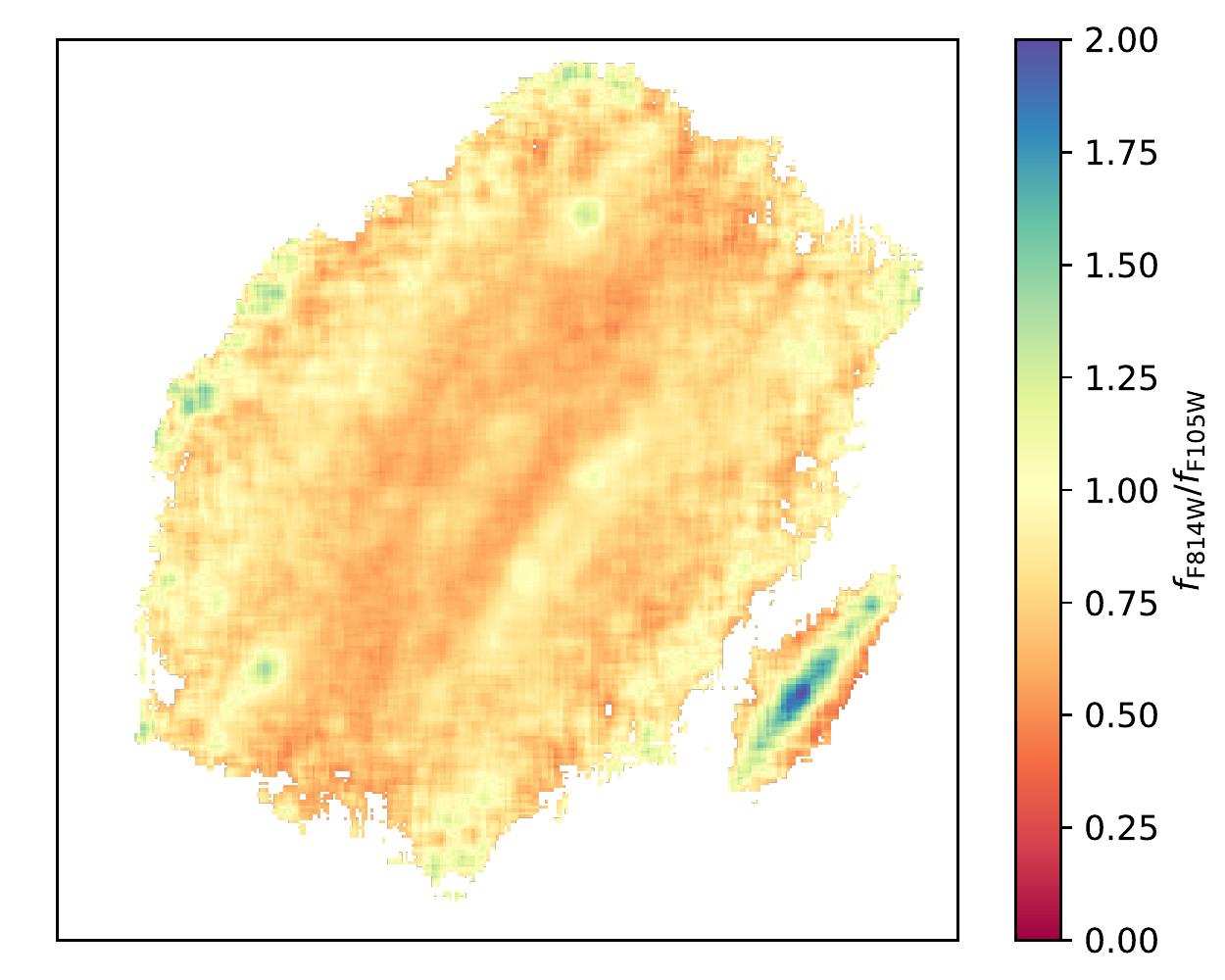}
    \caption{The ratio of flux in F814W to F105W filters after subtraction of the intracluster light. Slit F shown (marked in Figure \ref{fig:comparison}) has a very blue color compared to all other regions in the arc. In \S\ref{sec:src} we discuss a possible origin of the color.}
    \label{fig:color}
\end{figure}

The feature in slit F that exhibits both flux asymmetry in F814W and temporal variability in F105W and F140W is also anomalously blue compared to the rest of the arc. Figure \ref{fig:color} shows the F814W/F105W flux ratio after we subtract the ICL as described in \S\ref{sec:flux}. We warn here that the FFP epochs for F814W and F105W are separated by $\sim6$ months, so the anomalous color might be due to temporal variability over this time. However, a single visit in F105W (PI: Rodney, GO-13386), shown as the red point in the right panel of Figure \ref{fig:variability}, coincides with the FFP visits in F814W and has the same flux in slit F as the rest of F105W FFP data. Thus, we believe the anomalous blue color in slit F we detect from stacked images is caused by a real color variation and not temporal variabiltiy. This favors microlensing of a luminous source star that is bluer in color than the underlying mean stellar population of the source star-forming galaxy.

%%%%%%%%%%%%%%%%%%%%%%%%%%%%%%%%%%%%%%%%%%%%%%%%%%%%%%%%%%%%%%%%%%%%%%%%%%%%%%%
%%%%%%%%%%%%%%%%%%%%%%%%%%%%%%%%%%%%%%%%%%%%%%%%%%%%%%%%%%%%%%%%%%%%%%%%%%%%%%%

%%%%%%%%%%%%%%%%%%%%%%%%%%%%%%%%%%%%%%%%%%%%%%%%%%%%%%%%%%%%%%%%%%%%%%%%%%%%%%%
%%%%%%%%%%%%%%%%%%%%%%%%%%%%%%%%%%%%%%%%%%%%%%%%%%%%%%%%%%%%%%%%%%%%%%%%%%%%%%%

%%%%%%%%%%%%%%%%%%%%%%%%%%%%%%%%%%%%%%%%%%%%%%%%%%%%%%%%%%%%%%
\section{Microlensing simulations}
\label{sec:theory}
%%%%%%%%%%%%%%%%%%%%%%%%%%%%%%%%%%%%%%%%%%%%%%%%%%%%%%%%%%%%%%

We now present a microlensing simulation specific for the variable object detected in slit F.

The microlensing behavior of highly magnified stars depends on the local properties of the macro-lens model near the smooth critical curve. 
Macro-lens models are available from the Frontier Fields Lens Model project~\citep{2017ApJ...837...97L}. We adopt the model of \cite{2017A&A...600A..90C}, which includes this arc as a constraint. At the location of the arc under our investigation, and using the notation of \cite{2017ApJ...850...49V}, the total cluster surface mass density amounts to a convergence $\kappa_0= 0.66$. The local gradient of the magnification eigenvalue that cancels on the critical curve, $\bfd$, has a magnitude $|\bfd| \simeq 7\,{\rm arcmin}^{-1}$ and is nearly parallel to the principal axis of arc elongation.
%$\alpha \simeq 90^{\circ}$. 
These parameters can be approximated as uniform throughout the arc. The fold model~\citep{schneider1992gravitational} predicts that the magnification of each macro-image of a point source is $\mu \simeq 200\,(60\,{\rm mas}/\Delta\theta)$, where $\Delta\theta$ is the distance from each image to the macro-critical curve, and the fiducial value $\Delta\theta = 60\,{\rm mas}$ corresponds to half the separation between the pixel pairs in slit F showing flux asymmetry, temporal variability and color anomaly.

The macro-critical curve at the intersection with the arc is at a projected distance $\sim 25\,$kpc from the BCG. The study of \cite{Montes:2017yct} suggested that the local intracluster stellar population has a metallicity $[{\rm Fe}/{\rm H}] \approx 0.0$ and an age $\sim 2$--$3\,$Gyr, a result we confirm by modeling the surface brightness measurements in the seven HST filters using the stellar population synthesis code Flexible Stellar Population Synthesis (\texttt{FSPS})~\citep{conroy2009propagation, conroy2010propagation}. The surface brightness is normalized to $22.9\,{\rm mag}/{\rm arcsec}^2$ in F120W, from which we infer that intracluster stars make a local contribution to the lensing convergence of $\kappa_\star \approx 0.02$.

Microlensing by intracluster stars alters the smooth macro-critical curve into a corrugated network of micro-critical curves~\citep{2017ApJ...850...49V}, with a half width $\kappa_\star/|\bfd| \simeq 170\,$mas. We are interested in the hypothesis that the anomalies in slit F are due to the pair of macro-images of an underlying highly magnified star, each of which is $\Delta\theta \approx 60\,$mas from the macro critical curve. Applying the analytic results of \cite{2017ApJ...850...49V}, we find that the tentative image pair, one interior and one exterior to the macro-critical curve, have an expected rate of micro-caustic crossings of $1.0\,{\rm yr}^{-1}$ and $0.5\,{\rm yr}^{-1}$, respectively, times $(v_t/400\,{\rm km/s})$, where $v_t$ is the effective relative transverse peculiar velocity (see Eq.[12] of \cite{2017ApJ...850...49V}) between the lens cluster and the source galaxy. In the concordance cosmology, $v_t$ has a magnitude $\sim 400\,{\rm km/s}$ owing to the large-scale structure motions of the cluster and the source plus the motion of the Solar System relative to the cosmic rest frame.

%%%%%%%%%%%%%%%%%%%%%%%%%%%%%%%%%%%%%%%%%%%%%%%
\begin{figure*}[t]
    \centering \includegraphics[width=\columnwidth]{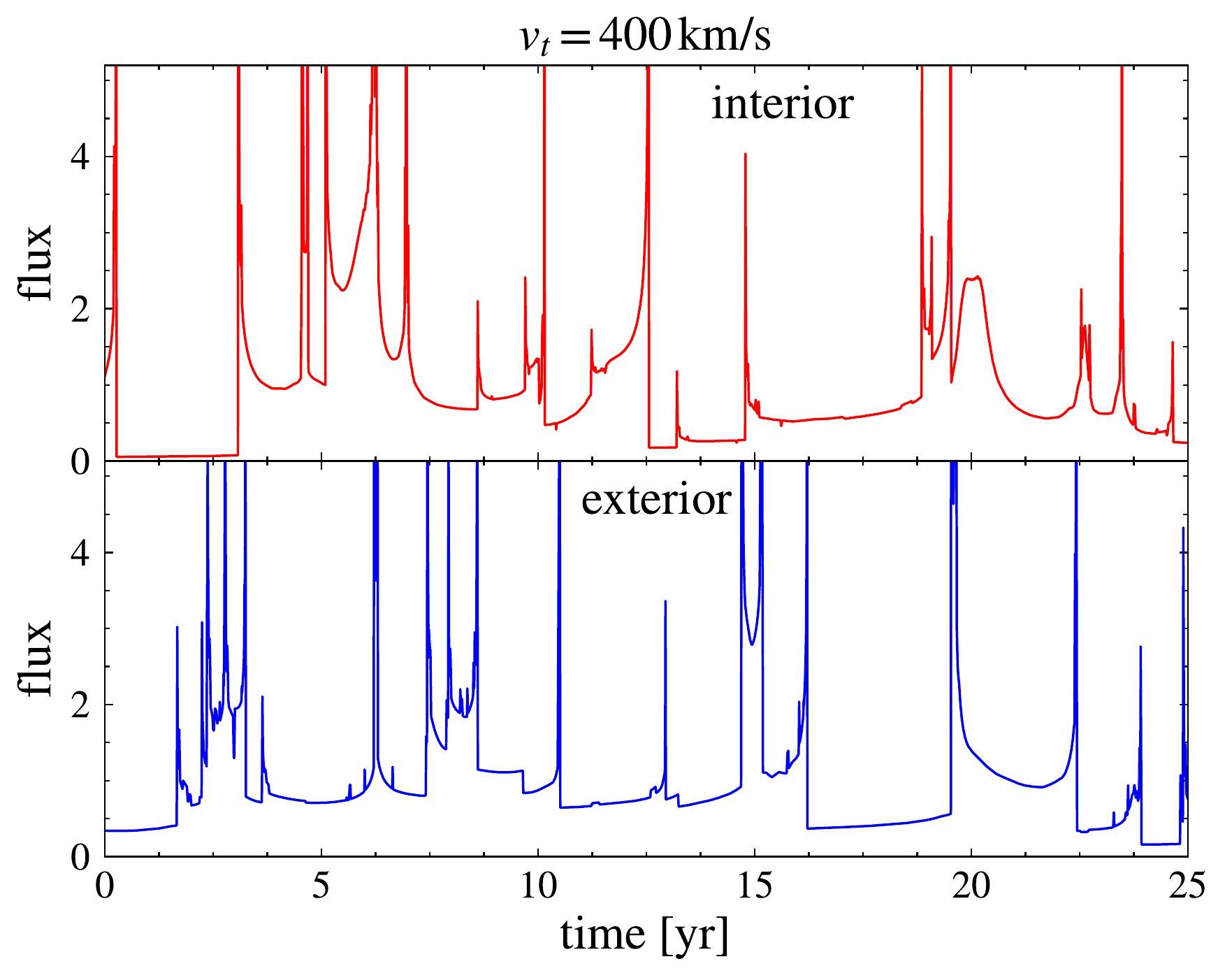}
    \includegraphics[width=\columnwidth]{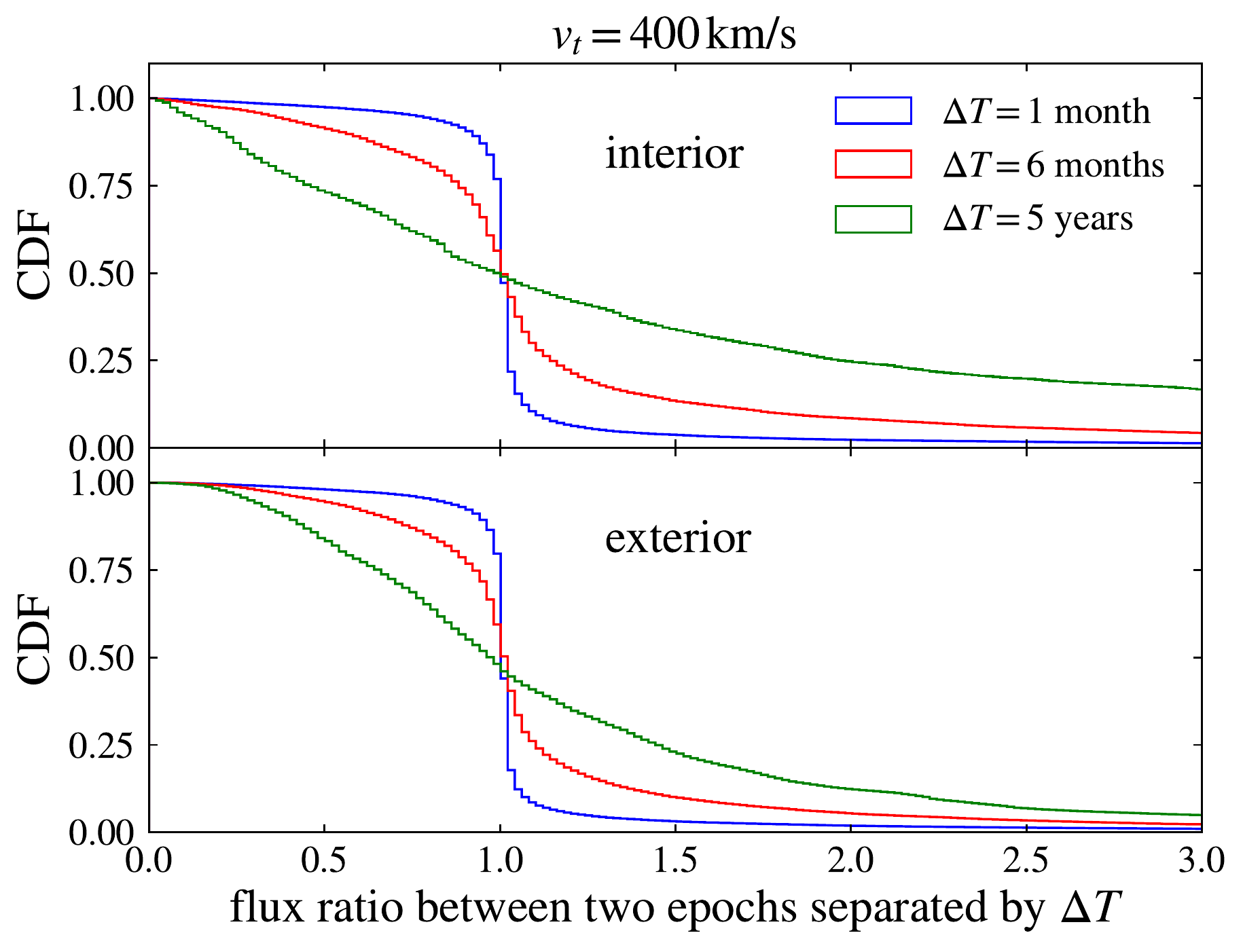}
    \caption{\textit{Left:} Example light curves for the pair of macro-images of a star (interior and exterior to the macro-critical line, for upper and lower panels, respectively) moving with a transverse velocity $v_t = 400\,\mathrm{km/s}$. The flux is normalized to the temporal mean. \textit{Right:} Cumulative probability distribution of the flux ratio between two observing epochs separated by 1 month, 6 months and 5 years for the interior (upper) and exterior images (lower panel). These typical light curves suggest that observations with a cadence from a few months to a few years will nearly always enable detection of flux variability.}
    \label{fig:lightcurve}
\end{figure*}
%%%%%%%%%%%%%%%%%%%%%%%%%%%%%%%%%%%%%%%%%%%%%%%

We simulate microlensing using a code of inverse ray-tracing, adopting the strategy of adaptive refinement as described in ~\cite{2018ApJ...867...24D}. We randomly sample micro-lens stars between $0.005\,M_\odot$ and $2\,M_\odot$ (the upper mass cutoff is appropriate for the aged intracluster stellar population) from the mass function of \cite{mattsson2010origin} and normalize the total mass to $\kappa_\star = 0.02$. We assume a source stellar radius $R=100\,R_\odot$, which is generally unresolved in the microlensing lightcurves except for the tip of the magnification peaks. In Figure \ref{fig:lightcurve}, we show sample light curves and the distribution of flux variability between two epochs separated by 1, 6 and 60 months, based on theoretical modeling of the target arc. We show results for both macro-images, which have different microlensing statistics. The results show that significant flux variability is unlikely over one month of observing baseline. This is consistent with the non-detection of variability between FFP visits, which spanned one month in the case of MACS J0416. As the observing baseline increases, the expected flux variations between grow. In particular, the $\sim$30\% flux variation that we measure between the two visits separated by about two years is likely.

We conclude that the variability observations are consistent with microlensing of a single, highly magnified star. This star is expected to undergo many micro-caustic crossings over many years, as shown in figure \ref{fig:lightcurve}. The crossing rate depends on several factors, including the exact separation between the two macro-images and the mass function of intracluster stars.

%%%%%%%%%%%%%%%%%%%%%%%%%%%%%%%%%%%%%%%%%%%%%%%%%%
\begin{figure}[h]
    \centering
    \includegraphics[width=\columnwidth]{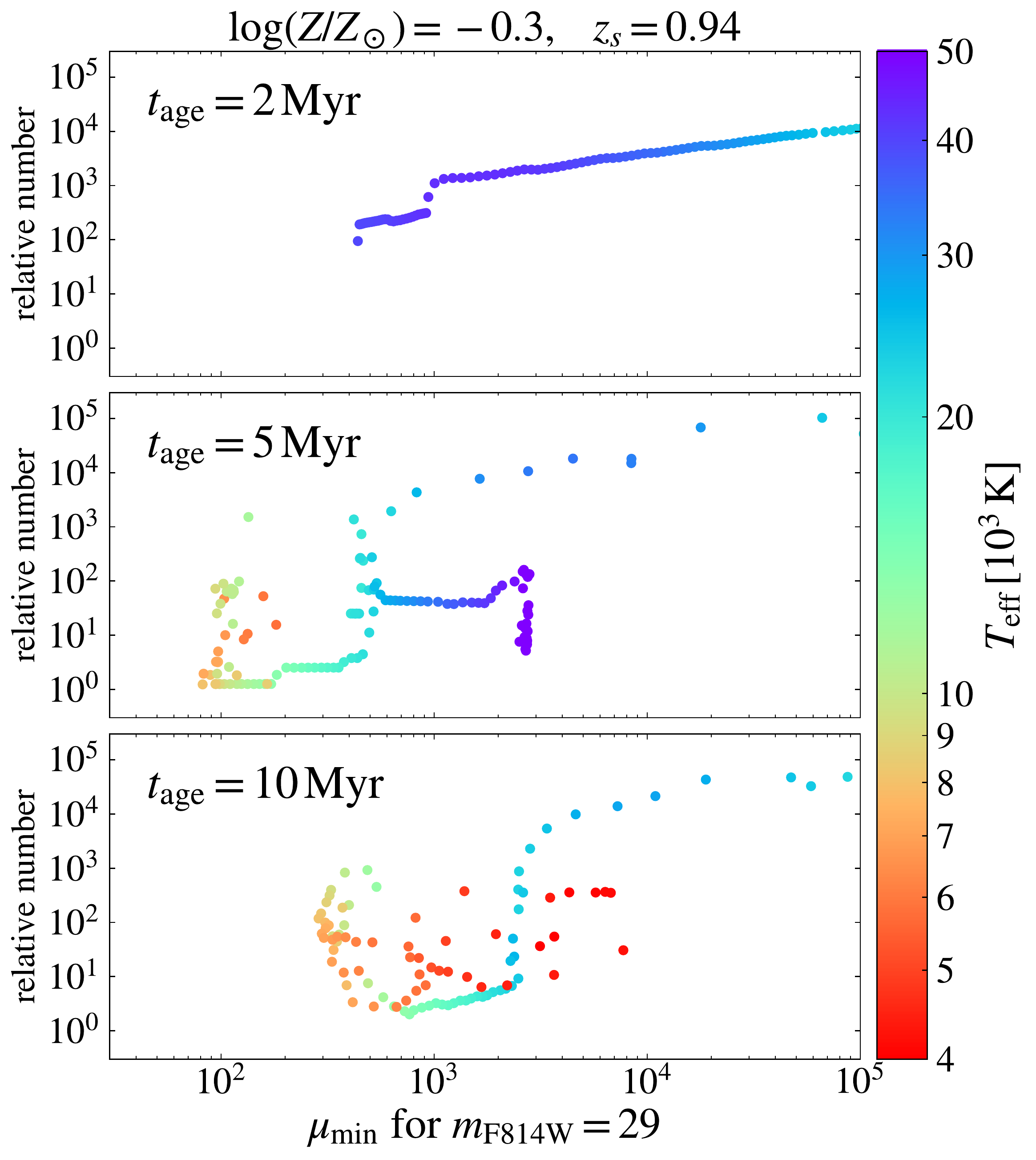}
    \caption{Minimum magnification factor (per macro-image) required for a star to have an observed magnitude $m=29$ (without dust attenuation) in F814W at $z_s=0.94$. Each point represents a type of star drawn from a single metallicity and age stellar population, with an abundance indicated by the vertical axis and an effective surface temperature $T_{\rm eff}$ shown by the color code. We consider three stellar ages $t_{\rm age}=2,\,5,\,10\,$Myr, and assume a metallicity $\log(Z/Z_\odot)=-0.3$, which best fits the spectral energy distribution of the star-forming complex in slit F. A 5 Myr-old stellar population contains hot main sequence stars detectable at $\mu \gtrsim 10^3$, and more luminous blue supergiants detectable at $\mu \gtrsim 10^2$. Main-sequence stars need to be fortuitously located much closer to micro-caustics to reach the threshold magnification factor.}
    \label{fig:ssp}
\end{figure}
%%%%%%%%%%%%%%%%%%%%%%%%%%%%%%%%%%%%%%%%%%%%%%%%%%

%%%%%%%%%%%%%%%%%%%%%%%%%%%%%%%%%%%%%%%%%%%%%%%%%%%%%%%%%%%%%%
\newpage
\section{Source star}
\label{sec:src}
%%%%%%%%%%%%%%%%%%%%%%%%%%%%%%%%%%%%%%%%%%%%%%%%%%%%%%%%%%%%%%

Explaining the anomalous asymmetry in slit F with microlensing of a source star
%the flux asymmetry in F814W during the FFP visits in 2014 
requires a flux difference between the two macro-images of $m_{\rm F814W} \simeq 29.5$. By comparison, a B-type supergiant with surface temperature $T_{\rm eff} \simeq 15000\,$K and radius $R \simeq 100\,R_\odot$ at $z=0.94$ has $m_{\rm F814W} \simeq 29.2 - 2.5\,\log_{10}(\mu/200)$ (without dust attenuation), where $\mu$ is the magnification factor per macro-image. From microlensing simulations, we find that the typical flux asymmetry between the image pair due to uncorrelated microlensing is similar to the mean flux. This suggests that an underlying blue supergiant (similar to the star detected in MACS J1149~\citep{2018NatAs...2..334K}) magnified by $\mu \sim 200$ near the caustic is a viable explanation for the observed flux asymmetry.

More generally, we study the type of star that is most likely to produce the asymmetry and variability we observe. We use \texttt{FSPS} to generate stellar evolution tracks including initial stellar masses up to $120\,M_\odot$. In Figure~\ref{fig:ssp}, we calculate the required minimal magnification factor per macro-image $\mu$ for a single star of each stellar type to reach $m=29$ in F814W at $z=0.94$. The abundance of each type of star is shown in the vertical axis. Assuming a metallicity $\log(Z/Z_\odot)=-0.3$ and the three indicated stellar ages, which fit the spectral energy distribution (SED) of the star-forming clump in slit F, we find that only supergiants are sufficiently bright if $\mu \sim 200$. These stars typically have $T_{\rm eff} < 15000\,$K and are only present if the stellar age is $t_{\rm age}=2$--$10\,$Myr. These conclusions remain largely valid for a range of metallicities $-1.5 < \log(Z/Z_\odot) < 0.0$.

Although red supergiants with $T_{\rm eff} < 6000\,$K may still be sufficiently bright in F814W, Figure~\ref{fig:ssp} shows that they are significantly rarer than B-type supergiants with $T_{\rm eff} = 10000$--$15000\,$K. We conclude that the most likely type of star that can reach the observed magnitude when highly magnified in a microlensing event is a blue supergiant. These stars spend a large fraction of their lifetime burning hydrogen on a shell around a helium core with a luminosity close to the Eddington value.

According to Figure~\ref{fig:ssp}, typical hot main-sequence stars are intrinsically fainter in F814W than supergiants. A main-sequence star may still explain the flux asymmetry if during the 2014 FFP visits the star temporarily acquired a microlensing-induced magnification factor (per macro-image) that is much larger than what the macro lens model predicts, $\mu \sim 200$. However, we disfavor this possibility as it has a low probability to occur at a randomly chosen epoch.

To reach a minimum magnification $\mu_{\rm min}$, a source star needs to be located within an area around micro-caustics that is $\propto 1/\mu^2_{\rm min}$. The abundance of lower luminosity stars therefore needs to be greater than that of higher luminosity ones by a factor $\propto \mu^2_{\rm min}$ to equally contribute to observed events at a fixed flux. This implies that caustic transiting stars are dominated by the most luminous stars as long as the luminosity function in a given band is $dN/dL \propto L^{-\alpha}$ with $\alpha < 3$. The bright end of the luminosity function measured from nearby star-forming galaxies, $\alpha \approx 2.5$, indeed falls into this regime~\citep{bresolin1998hubble}, while in regions of vigorous star formation the slope is even shallower $\alpha \lesssim 2$~\citep{malumuth1994ubv}. This predicts that the brightest highly magnified stars should be supergiants.

Whether a blue supergiant can precisely explain the observed color anomaly between F814W and F105W is uncertain. Extracting the star SED taking into account the PSF is complicated due to blending of the highly magnified star with light from any surrounding star-forming region, likely associated with the star. We leave this for future analysis.

%%%%%%%%%%%%%%%%%%%%%%%%%%%%%%%%%%%%%%%%%%%%%%%%%%
\section{Statistical analysis of flux asymmetry}
\label{sec:fits}
%%%%%%%%%%%%%%%%%%%%%%%%%%%%%%%%%%%%%%%%%%%%%%%%%%%%%%%%%%%%%%%%%%%%%%%%%%%%%%%%%

Flux asymmetries in arcs caused by stellar microlensing or by DM substructure should be more frequent closer to the critical curve (see \S\ref{sec:theory} and \cite{2018ApJ...867...24D}). In \S\ref{sec:flux}, we detected individual flux asymmetries in specific position pairs. An alternative approach is to use the flux data for the whole arc and statistically test the symmetry of the two-dimensional surface brightness pattern, to check if asymmetry increases close to the critical curve. This idea is along the lines of detecting substructure in galaxy lenses from residuals in fitting a smooth lens model to the observed surface brightness pattern~\citep{hezaveh2013dark, asadi2017probing, birrer2017lensing}.
%The approach requires solving for the local lens model. 
In the subsections below we describe our method, and demonstrate how it provides additional evidence for flux asymmetry.

Throughout this section we model only the arc structure containing slits A through E. We exclude slit F from consideration, because it contains the clearly detected individual asymmetry discussed above and is at a large separation from the rest of the arc, which complicates our lens model fitting procedure.

%%%%%%%%%%%%%%%%%%%%%%%%%%%%%%%%%%%%%%%%%%%%%%%%%%%%%%%%%%%%%%%%%%%%%%%%%%%%%%%%%
\subsection{Fitting the lens model}
\label{sec:fold}
%%%%%%%%%%%%%%%%%%%%%%%%%%%%%%%%%%%%%%%%%%%%%%%%%%

%%%%%%%%%%%%%%%%%%%%%%%%%%%%%%%%%%%%%%%%%%%%%%%%%%%%%%%%%%%%%%%%%%%%%%%%%%%%%%%%%
\begin{figure}
    \centering
    \includegraphics[width=\columnwidth]{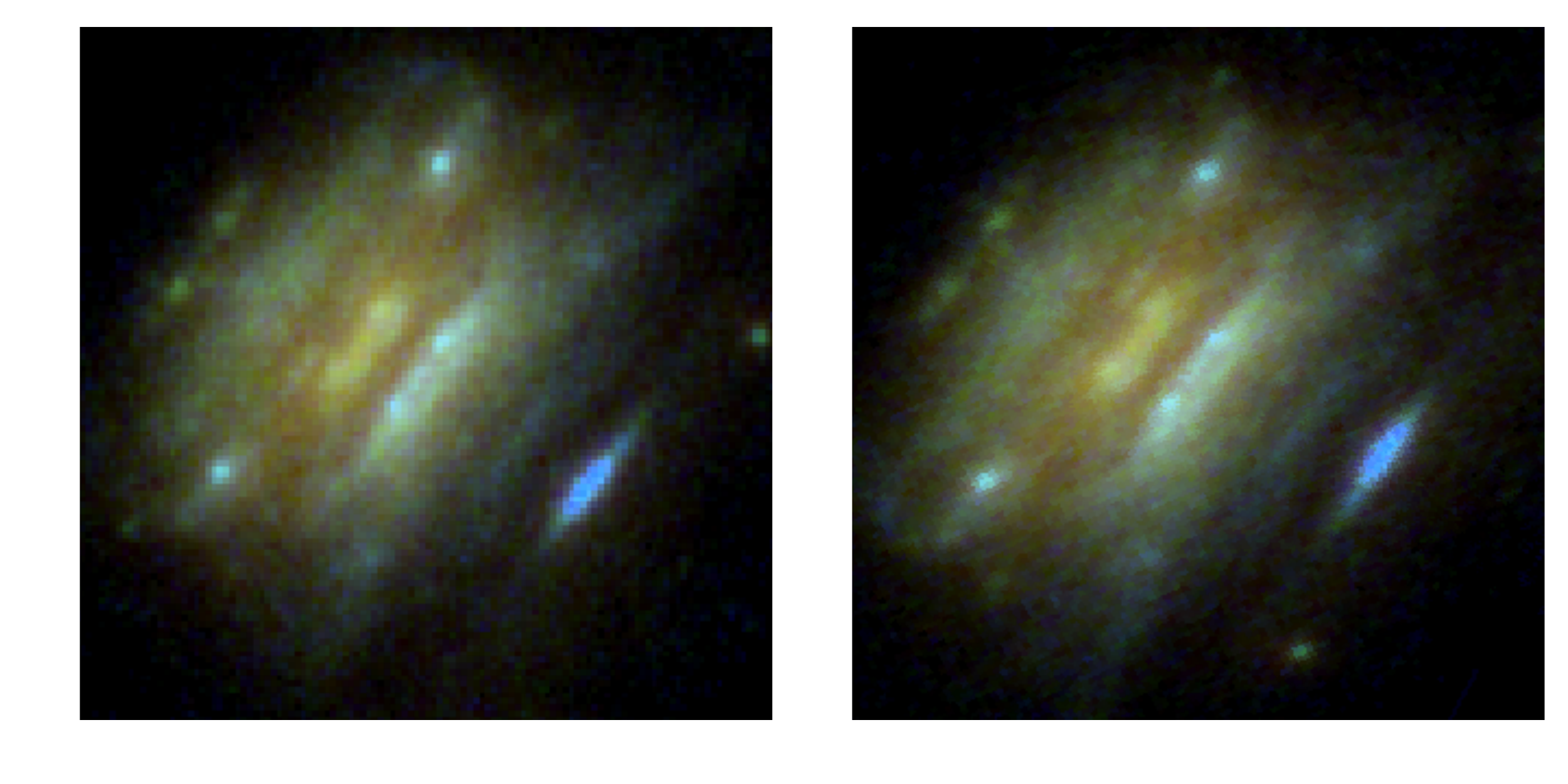}
    \caption{Combined false-color cutout ($3.5\arcsec\times3.5\arcsec$) using F105W, F814W and F606W showing the arc (left) and its flipped image using the second-order best-fit model (right) as described in \S\ref{sec:fold}. The blue feature in the right bottom corner was not used in the fit, so its position shifts significantly. By visually comparing these two images one can asses the asymmetry of the images.}
    \label{fig:flip}
\end{figure}
%%%%%%%%%%%%%%%%%%%%%%%%%%%%%%%%%%%%%%%%%%%%%%%%%%%%%%%%%%%%%%%%%%%%%%%%%%%%%%%%%

We start by fitting the second-order fold model \citep[e.g.][]{1992grle.book.....S} to the F814W image. We do not attempt to use other filters for fitting for the reasons described in \S\ref{sec:flux}. The fitting procedure minimizes the mismatch between the original image and the image flipped about the critical curve. There are multiple approaches to this optimization problem.

%%%%%%%%%%%%%%%%%%%%%%%%%%%%%%%%%%%%%%%%%%%%%%%%%%%%%%%%%%%%%%%%%%%%%%%%%%%%%%%%%
\begin{figure*}
    \centering
    \includegraphics[width=\textwidth]{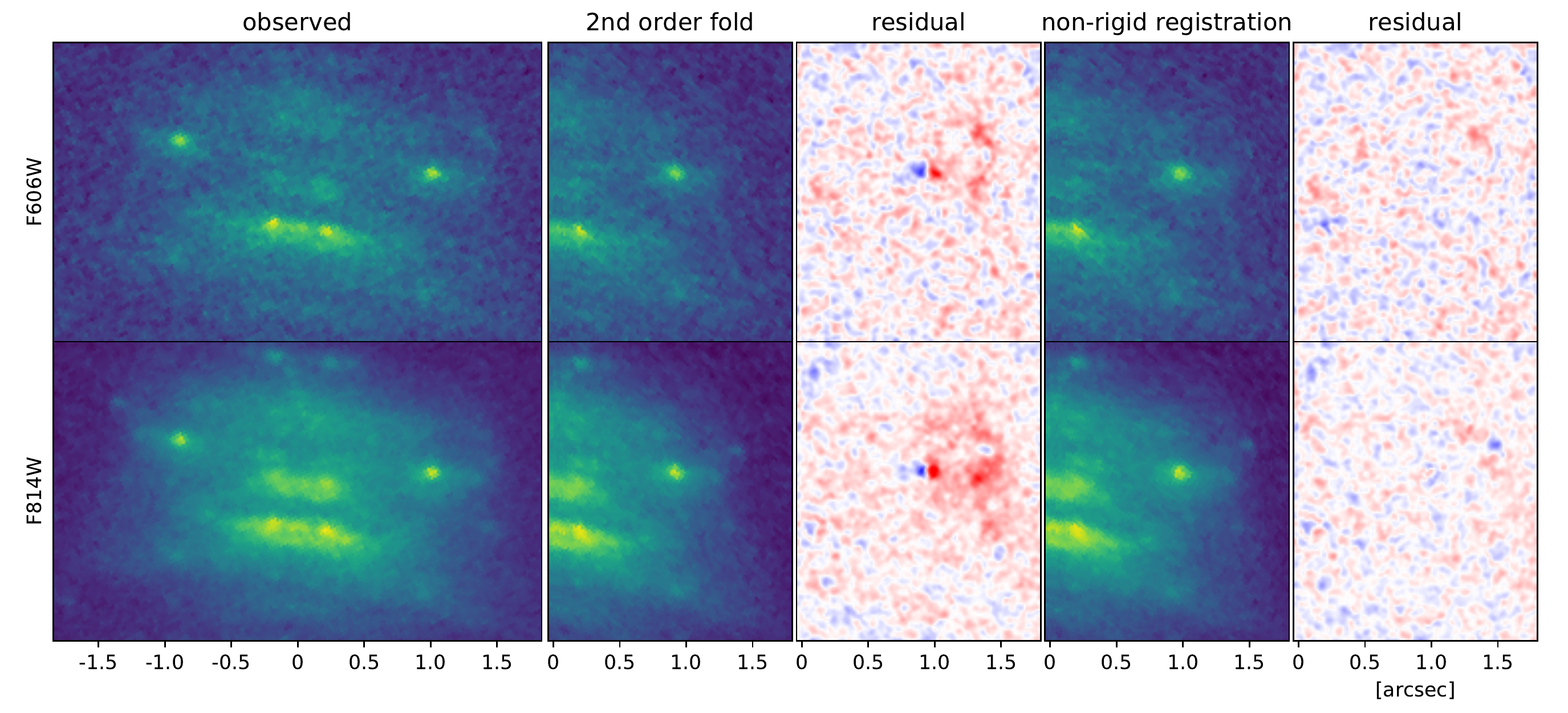}
    \caption{\textit{First column:} Image cutouts of $1.8\arcsec\times2.6\arcsec$ showing the arc in F606W and F814W, rotated to have the critical curve (which is not a perfect straight line) largely aligned with the vertical axis. \textit{Second and third columns:} flipped left image of the arc following the second-order best-fit fold model as described in \S\ref{sec:fold}, and its mismatch (or ``residual'') with the original right image measured in units of the noise. \textit{Fourth and fifth columns:} images adjusted with the non-rigid transformation discussed in \S\ref{sec:fold}, and corresponding residuals. Residuals are decreased by the adjustment. The color scales used for the third and fifth columns are the same.}
    \label{fig:cutouts}
\end{figure*}
%%%%%%%%%%%%%%%%%%%%%%%%%%%%%%%%%%%%%%%%%%%%%%%%%%%%%%%%%%%%%%%%%%%%%%%%%%%%%%%%%

We decided to fit only data within a $0.5\arcsec$ vicinity of the expected position of the critical curve in the lens model we use. Including the entire arc substantially worsens the fit near the critical curve, because the departure from a simple fold model increases as data from more distant pixels is included. We are also interested in regions close to the critical curve, where effects of microlensing and subhalo lensing are enhanced. The result of this fit is shown in Figure \ref{fig:flip} and in the second and third columns of Figure \ref{fig:cutouts}. The image in F606W is transformed using the parameters we derive from the F814W image.

The residual map derived using the second order fold model is shown in the third column of Figure \ref{fig:cutouts}. The map shows systematic deviations indicating that even the best fit model cannot describe the local lens model. Thus, we explore the possibility of adjusting the lens model using non-rigid image registration, a method that finds a continuous transformation from one image to another one that is similar. In our case, we transform between the original and `flipped' images of the arc in F814W. We adopt the method of generic diffeomorphic groupwise registration (GDGRegistration) as implementated in the Python package \texttt{pirt}\footnote{\url{https://pypi.org/project/pirt/}}. We apply the same transformation to F606W.

Non-rigid image registration has some disadvantages for our problem: it is not parameter free, and the solution may not be unique. %These are the common issues with many other algorithms for finding substructure in the gravitational lenses.
The transformation generated by the algorithm may also have rotation, i.e., it may not be derived from the gradient of a scalar lensing potential (which may occur in real situations involving multiple lens planes), but they are constrained to conserve surface brightness and not to create new caustics. The main parameters controlling how the fitting procedure is done are the smallest allowed scale of deformation and its maximum amplitude. When extreme values for these parameters are taken, the algorithm can effectively erase any feature by collapsing the region into a point. We tune these parameters to obtain the expected noise level in the residual map ($\chi^2/\rm{DOF} \sim 1$) fit over the whole arc. The result of this fitting procedure is shown in the last two columns of Figure \ref{fig:cutouts}. 

%%%%%%%%%%%%%%%%%%%%%%%%%%%%%%%%%%%%%%%%%%%%%%%%%%%%%%%%%%%%%%%%%%%%%%%%%%%%%%%%%
\subsection{Results}
\label{sec:lensresult}
%%%%%%%%%%%%%%%%%%%%%%%%%%%%%%%%%%%%%%%%%%%%%%%%%%%%%%%%%%%%%%%%%%%%%%%%%%%%%%%%%

\begin{figure}
    \centering
    \includegraphics[width=\columnwidth]{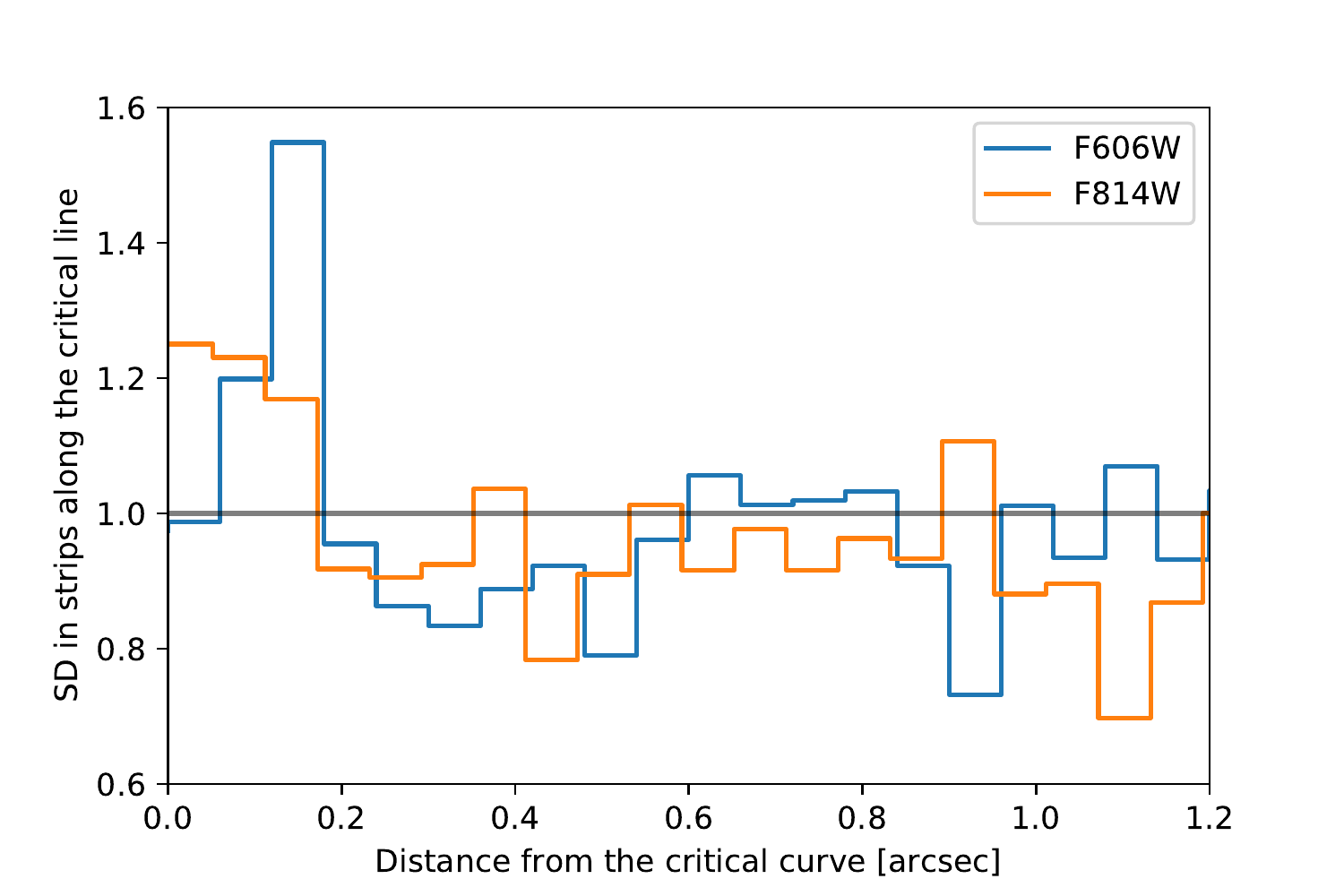}
    \caption{Standard deviation of residuals in the F814W and F606W images of the arc, excluding slit F, as a function of distance from the critical curve, using the lens model described in \S\ref{sec:fits}. The standard deviation is calculated in $\sim0.06\arcsec$ wide strips parallel to the critical curve using the residuals in the rightmost column in Figure \ref{fig:cutouts}. The peak in the standard deviation within $\sim0.2\arcsec$ may be caused by differential microlensing of stars or star-forming regions that are not individually detectable.
    %See \S\ref{sec:lensresult} for more details.
    }
    \label{fig:deviations}
\end{figure}

In Figure \ref{fig:deviations} we show the standard deviation of flux residuals on the two sides of the critical line evaluated from the rightmost column in Figure \ref{fig:cutouts}, along $0.06\arcsec$ wide vertical strips. An increased dispersion within $0.2\arcsec$ is seen in both F606W and F814W filters. This may be interpreted as the asymmetry caused by several microlensed stars that cannot be individually detected, but we cannot robustly detect the effect without additional data and careful analysis. 
Several caveats limit our ability to properly estimate the significance of the peak in dispersion close to the critical line, including the complexity and large number of free parameters of our analysis.

In the future, many microlensed stars and star-forming structures can be analyzed in this way, evaluating the required lensing perturbers to explain the degree of asymmetry in deeper and higher-resolution images produced by the next generation observatories.

%%%%%%%%%%%%%%%%%%%%%%%%%%%%%%%%%%%%%%%%%%%%%%%%%%%%%%%%%%%%%%%%
%%%%%%%%%%%%%%%%%%%%%%%%%%%%%%%%%%%%%%%%%%%%%%%%%%%%%%%%%%%%%%%%

\newpage

%%%%%%%%%%%%%%%%%%%%%%%%%%%%%%%%%%%%%%%%%%%%%%%%%%%%%%%%%%%%%%%%%%%%%%%%%%%%%%%%
\section{Discussion}
\label{sec:discussion}
%%%%%%%%%%%%%%%%%%%%%%%%%%%%%%%%%%%%%%%%%%%%%%%%%%%%%%%%%%%%%%%%%%%%%%%%%%%%%%%%

We have searched for transients in a caustic-straddling lensed arc in the cluster lensing system MACS J0416. Detecting individual highly magnified stars is promising in this arc thanks to the relatively low arc redshift, $z \simeq 0.94$, and the existence of star-forming complexes hosting highly luminous stars. The large angular size of the arc, $\sim 4\arcsec \times 4\arcsec$, is advantageous for applying astrometric methods to probe small-scale lensing perturbations from DM substructure to the critical curve~\citep{2018ApJ...867...24D}.  

The main finding of this paper is asymmetric structures in the arc seen in multiple slits along the critical curve in F606W and F814W filters, and temporal flux variability in slit F in F105W between the 2012 visits and the 2014 FFP visits. We believe that these observations require variability of a highly magnified star induced by microlensing at least in slit F. Additional microlensed stars may be present in other parts of the arc.  

Our microlensing simulations show that images of highly magnified stars inevitably vary in flux over time baselines of years, due to the ``floor'' in lightcurves jumping between microlensing peaks (see Figure~\ref{fig:lightcurve}). The brief microlensing peaks should also be detectable even though they are infrequent.

Multiband NIRCam images ($0.8$--$5.0\,\mu$m) of MACS J0416 will be obtained with the scheduled JWST GTO program (PI: Windhorst, \#1176)\footnote{\url{https://jwst.stsci.edu/observing-programs/program-information?id=1176}}. The cluster will be visited at three epochs in Cycle 1 separated by $\sim$few days, 180 days, and 360 days. We suggest that flux variability can be verified at the various spots of surface brightness asymmetry through a comparison between various JWST visits and the existing HST visits. At infrared wavelengths $\lambda \gtrsim 1.6\,\mu$m, JWST may also uncover highly magnified red supergiants which are not as bright in HST filters.

The methods we have adopted are applicable to the search of caustic transients in other caustic-straddling arcs. The chance of detecting microlensing flux asymmetry and temporal variability depends on data quality, observing cadence, as well as lens and source properties in individual systems, but in general microlensed stars can be identified with deep observations at a few epochs without requiring detailed monitoring. To gather useful information about the macro-lens model, arc stellar population, and the intracluster stellar population, follow-up imaging and spectroscopic studies of caustic-straddling arcs and their vicinity are highly desirable.

\vspace{10pt}
As mentioned in the introduction, toward the completion of this work a preprint \cite{Chen:2019ncy} appeared in which the authors analyzed HST images of MACS J0416 and concluded that a blue super-giant was detected exactly at the location of the flux asymmetry in our slit F. The detection is strongly supported by a bright microlensing event (reached $m_{\rm F814W}=26.4$) in the light curve that took place in Semptember of 2014 shortly after the FFP visits we use. In this work, we have done analysis independently of their work, and have detected a brightening event during the 2012 visits compared to the 2014 FFP visits, which corresponds to flux change at a much weaker level. We examined the extra visits that were used by \cite{Chen:2019ncy} but are not used here, and confirmed the brightening event in slit F. In addition, we looked into the more recent F606W and F814W visits in 2019 (PI:Steinhardt GO-15117) and did not detect any variability compared to the 2014 FFP visits.

Our study suggests that the source star probably had a magnitude around $m_{\rm F814W} \simeq 29$ at the 2014 FFP epochs around the floor of its microlensing light curve. The flux change discussed in this paper likely reflects jumps in the ``floor'' magnitude across microlensing peaks, which we predict to be inevitable when comparing flux measurements separated by years, while the event discussed in \cite{Chen:2019ncy} is most likely caused by the same star undergoing a microlensing peak resembling those in Figure \ref{fig:lightcurve}. Our conclusions about the nature of the caustic anomaly and the property of the magnified star are consistent with those of \cite{Chen:2019ncy}.

% \vspace{20pt}
%------------------------------------------------------------------------------
%%%%%%%%%%%%%%%%%%%%%%%%%%%%%%%%%%%%%%%%%%%%%%%%
\begin{acknowledgments}

AAK is supported by the William D. Loughlin Membership Fund. LD acknowledges the support from the Raymond and Beverly Sackler Foundation Fund. TV acknowledges support by the Friends of the Institute for Advanced Study. JM is supported by Spanish fellowship PRX18/00444, and by the Corning Glass Works Foundation Fellowship Fund.

\end{acknowledgments}
%%%%%%%%%%%%%%%%%%%%%%%%%%%%%%%%%%%%%%%%%%%%%%%%
% \newpage
\bibliography{refs}

\begin{thebibliography}{}
\expandafter\ifx\csname natexlab\endcsname\relax\def\natexlab#1{#1}\fi
\providecommand{\url}[1]{\href{#1}{#1}}

\bibitem[{Alcock {et~al.}(2001)Alcock, Allsman, Alves, Axelrod, Becker,
  Bennett, Cook, Dalal, Drake, Freeman, {et~al.}}]{alcock2001macho}
Alcock, C., Allsman, R., Alves, D.~R., {et~al.} 2001, The Astrophysical Journal
  Letters, 550, L169

\bibitem[{Asadi {et~al.}(2017)Asadi, Zackrisson, \&
  Freeland}]{asadi2017probing}
Asadi, S., Zackrisson, E., \& Freeland, E. 2017, Monthly Notices of the Royal
  Astronomical Society, 472, 129

\bibitem[{Birrer {et~al.}(2017)Birrer, Amara, \& Refregier}]{birrer2017lensing}
Birrer, S., Amara, A., \& Refregier, A. 2017, Journal of Cosmology and
  Astroparticle Physics, 2017, 037

\bibitem[{Bresolin {et~al.}(1998)Bresolin, Kennicutt~Jr, Ferrarese, Gibson,
  Graham, Macri, Phelps, Rawson, Sakai, Silbermann,
  {et~al.}}]{bresolin1998hubble}
Bresolin, F., Kennicutt~Jr, R.~C., Ferrarese, L., {et~al.} 1998, The
  Astronomical Journal, 116, 119

\bibitem[{{Caminha} {et~al.}(2017){Caminha}, {Grillo}, {Rosati}, {Balestra},
  {Mercurio}, {Vanzella}, {Biviano}, {Caputi}, {Delgado-Correal}, {Karman},
  {Lombardi}, {Meneghetti}, {Sartoris}, \& {Tozzi}}]{2017A&A...600A..90C}
{Caminha}, G.~B., {Grillo}, C., {Rosati}, P., {et~al.} 2017, \aap, 600, A90

\bibitem[{Chen {et~al.}(2019)}]{Chen:2019ncy}
Chen, W., {et~al.} 2019, arXiv:1902.05510

\bibitem[{Conroy \& Gunn(2010)}]{conroy2010propagation}
Conroy, C., \& Gunn, J.~E. 2010, The Astrophysical Journal, 712, 833

\bibitem[{Conroy {et~al.}(2009)Conroy, Gunn, \& White}]{conroy2009propagation}
Conroy, C., Gunn, J.~E., \& White, M. 2009, The Astrophysical Journal, 699, 486

\bibitem[{{Dai} {et~al.}(2018){Dai}, {Venumadhav}, {Kaurov}, \&
  {Miralda-Escud}}]{2018ApJ...867...24D}
{Dai}, L., {Venumadhav}, T., {Kaurov}, A.~A., \& {Miralda-Escud}, J. 2018,
  \apj, 867, 24

\bibitem[{Diego(2018)}]{diego2018universe}
Diego, J.~M. 2018, arXiv preprint arXiv:1806.04668

\bibitem[{{Diego} {et~al.}(2018){Diego}, {Kaiser}, {Broadhurst}, {Kelly},
  {Rodney}, {Morishita}, {Oguri}, {Ross}, {Zitrin}, {Jauzac}, {Richard},
  {Williams}, {Vega-Ferrero}, {Frye}, \& {Filippenko}}]{2018ApJ...857...25D}
{Diego}, J.~M., {Kaiser}, N., {Broadhurst}, T., {et~al.} 2018, \apj, 857, 25

\bibitem[{Griest {et~al.}(2013)Griest, Cieplak, \& Lehner}]{griest2013new}
Griest, K., Cieplak, A.~M., \& Lehner, M.~J. 2013, Physical review letters,
  111, 181302

\bibitem[{Hezaveh {et~al.}(2013)Hezaveh, Dalal, Holder, Kuhlen, Marrone,
  Murray, \& Vieira}]{hezaveh2013dark}
Hezaveh, Y., Dalal, N., Holder, G., {et~al.} 2013, The Astrophysical Journal,
  767, 9

\bibitem[{{Kelly} {et~al.}(2018){Kelly}, {Diego}, {Rodney}, {Kaiser},
  {Broadhurst}, {Zitrin}, {Treu}, {P{\'e}rez-Gonz{\'a}lez}, {Morishita},
  {Jauzac}, {Selsing}, {Oguri}, {Pueyo}, {Ross}, {Filippenko}, {Smith},
  {Hjorth}, {Cenko}, {Wang}, {Howell}, {Richard}, {Frye}, {Jha}, {Foley},
  {Norman}, {Bradac}, {Zheng}, {Brammer}, {Benito}, {Cava}, {Christensen}, {de
  Mink}, {Graur}, {Grillo}, {Kawamata}, {Kneib}, {Matheson}, {McCully},
  {Nonino}, {P{\'e}rez-Fournon}, {Riess}, {Rosati}, {Schmidt}, {Sharon}, \&
  {Weiner}}]{2018NatAs...2..334K}
{Kelly}, P.~L., {Diego}, J.~M., {Rodney}, S., {et~al.} 2018, Nature Astronomy,
  2, 334

\bibitem[{{Lotz} {et~al.}(2017){Lotz}, {Koekemoer}, {Coe}, {Grogin}, {Capak},
  {Mack}, {Anderson}, {Avila}, {Barker}, {Borncamp}, {Brammer}, {Durbin},
  {Gunning}, {Hilbert}, {Jenkner}, {Khandrika}, {Levay}, {Lucas}, {MacKenty},
  {Ogaz}, {Porterfield}, {Reid}, {Robberto}, {Royle}, {Smith},
  {Storrie-Lombardi}, {Sunnquist}, {Surace}, {Taylor}, {Williams}, {Bullock},
  {Dickinson}, {Finkelstein}, {Natarajan}, {Richard}, {Robertson}, {Tumlinson},
  {Zitrin}, {Flanagan}, {Sembach}, {Soifer}, \&
  {Mountain}}]{2017ApJ...837...97L}
{Lotz}, J.~M., {Koekemoer}, A., {Coe}, D., {et~al.} 2017, \apj, 837, 97

\bibitem[{Malumuth \& Heap(1994)}]{malumuth1994ubv}
Malumuth, E.~M., \& Heap, S.~R. 1994, The Astronomical Journal, 107, 1054

\bibitem[{Mattsson(2010)}]{mattsson2010origin}
Mattsson, L. 2010, Astronomy \& Astrophysics, 515, A68

\bibitem[{Minor {et~al.}(2017)Minor, Kaplinghat, \& Li}]{minor2017robust}
Minor, Q.~E., Kaplinghat, M., \& Li, N. 2017, The Astrophysical Journal, 845,
  118

\bibitem[{{Miralda-Escude}(1991)}]{1991ApJ...379...94M}
{Miralda-Escude}, J. 1991, \apj, 379, 94

\bibitem[{Montes \& Trujillo(2018)}]{Montes:2017yct}
Montes, M., \& Trujillo, I. 2018, Mon. Not. Roy. Astron. Soc., 474, 917

\bibitem[{Niikura {et~al.}(2017)Niikura, Takada, Yasuda, Lupton, Sumi, More,
  More, Oguri, \& Chiba}]{niikura2017microlensing}
Niikura, H., Takada, M., Yasuda, N., {et~al.} 2017, arXiv preprint
  arXiv:1701.02151

\bibitem[{Oguri {et~al.}(2018)Oguri, Diego, Kaiser, Kelly, \&
  Broadhurst}]{Oguri:2017ock}
Oguri, M., Diego, J.~M., Kaiser, N., Kelly, P.~L., \& Broadhurst, T. 2018,
  Phys. Rev., D97, 023518

\bibitem[{{Rodney} {et~al.}(2018){Rodney}, {Balestra}, {Bradac}, {Brammer},
  {Broadhurst}, {Caminha}, {Chiriv{\i}}, {Diego}, {Filippenko}, {Foley},
  {Graur}, {Grillo}, {Hemmati}, {Hjorth}, {Hoag}, {Jauzac}, {Jha}, {Kawamata},
  {Kelly}, {McCully}, {Mobasher}, {Molino}, {Oguri}, {Richard}, {Riess},
  {Rosati}, {Schmidt}, {Selsing}, {Sharon}, {Strolger}, {Suyu}, {Treu},
  {Weiner}, {Williams}, \& {Zitrin}}]{2018NatAs...2..324R}
{Rodney}, S.~A., {Balestra}, I., {Bradac}, M., {et~al.} 2018, Nature Astronomy,
  2, 324

\bibitem[{Schneider(1992)}]{schneider1992gravitational}
Schneider, P. 1992, in Gravitational Lenses (Springer), 196--208

\bibitem[{{Schneider} {et~al.}(1992){Schneider}, {Ehlers}, \&
  {Falco}}]{1992grle.book.....S}
{Schneider}, P., {Ehlers}, J., \& {Falco}, E.~E. 1992, {Gravitational Lenses},
  112, doi:10.1007/978-3-662-03758-4

\bibitem[{Tisserand {et~al.}(2007)Tisserand, Le~Guillou, Afonso, Albert,
  Andersen, Ansari, Aubourg, Bareyre, Beaulieu, Charlot,
  {et~al.}}]{tisserand2007limits}
Tisserand, P., Le~Guillou, L., Afonso, C., {et~al.} 2007, Astronomy \&
  Astrophysics, 469, 387

\bibitem[{{Venumadhav} {et~al.}(2017){Venumadhav}, {Dai}, \&
  {Miralda-Escud{\'e}}}]{2017ApJ...850...49V}
{Venumadhav}, T., {Dai}, L., \& {Miralda-Escud{\'e}}, J. 2017, \apj, 850, 49

\bibitem[{Windhorst {et~al.}(2018)Windhorst, Timmes, Wyithe, Alpaslan, Andrews,
  Coe, Diego, Dijkstra, Driver, Kelly, {et~al.}}]{windhorst2018observability}
Windhorst, R.~A., Timmes, F., Wyithe, J. S.~B., {et~al.} 2018, The
  Astrophysical Journal Supplement Series, 234, 41

\bibitem[{{Zitrin} {et~al.}(2013){Zitrin}, {Meneghetti}, {Umetsu},
  {Broadhurst}, {Bartelmann}, {Bouwens}, {Bradley}, {Carrasco}, {Coe}, {Ford},
  {Kelson}, {Koekemoer}, {Medezinski}, {Moustakas}, {Moustakas}, {Nonino},
  {Postman}, {Rosati}, {Seidel}, {Seitz}, {Sendra}, {Shu}, {Vega}, \&
  {Zheng}}]{2013ApJ...762L..30Z}
{Zitrin}, A., {Meneghetti}, M., {Umetsu}, K., {et~al.} 2013, \apj, 762, L30

\end{thebibliography}
\end{document}